\documentclass[prd,tightenlines,nofootinbib,superscriptaddress]{revtex4}

\usepackage{amsfonts,amssymb,amsthm,bbm}

\usepackage{amsmath}

\usepackage{hyperref}

\usepackage{subcaption}
\usepackage{color,psfrag}
\usepackage[dvips]{graphicx}

\usepackage{tikz}
\usetikzlibrary{calc}
\usetikzlibrary{decorations.pathmorphing}
\usetikzlibrary{shapes.geometric}
\usetikzlibrary{arrows,decorations.markings}
\usetikzlibrary{patterns}

\newcommand{\C}{{\mathbb C}}
\newcommand{\N}{{\mathbb N}}
\newcommand{\R}{{\mathbb R}}

\newcommand{\cA}{{\mathcal A}}

\newcommand{\cF}{{\mathcal F}}

\newcommand{\cH}{{\mathcal H}}

\newcommand{\cN}{{\mathcal N}}

\newcommand{\cP}{{\mathcal P}}

\newcommand{\cT}{{\mathcal T}}
\newcommand{\cV}{{\mathcal V}}
\newcommand{\cD}{{\mathcal D}}
\newcommand{\cC}{{\mathcal C}}

\newcommand{\cS}{{\mathcal S}}

\newcommand{\cI}{{\mathcal I}}

\newcommand{\SU}{\mathrm{SU}}

\newcommand{\be}{\begin{equation}}
\newcommand{\ee}{\end{equation}}
\newcommand{\beq}{\begin{eqnarray}}
\newcommand{\eeq}{\end{eqnarray}}
\newcommand{\bes}{\begin{eqnarray}}
\newcommand{\ees}{\end{eqnarray}}

\renewcommand{\u}{{\mathfrak{u}}}
\newcommand{\su}{{\mathfrak{su}}}

\newcommand{\la}{\langle}
\newcommand{\ra}{\rangle}

\newcommand{\tr}{{\mathrm{Tr}}}
\newcommand{\f}{\frac}

\def\nn{\nonumber}
\def\pp{\partial}

\def\eps{\epsilon}

\newcommand{\id}{\mathbb{I}}

\def\rd{\textrm{d}}
\def\tpsi{\tilde{\psi}}
\def\vJ{\vec{J}}
\def\tj{\tilde{j}}
\def\tm{\tilde{m}}
\def\tn{\tilde{n}}
\def\ua{\uparrow}
\def\da{\downarrow}

\newcommand{\blue}{\color{blue}}
\newcommand{\red}{\color{red}}
\newcommand{\green}{\color{green}}
\newcommand{\orange}{\color{orange}}

\def\centerarc[#1](#2)(#3:#4:#5)
{ \draw[#1] ($(#2)+({#5*cos(#3)},{#5*sin(#3)})$) arc (#3:#4:#5); }

\def\centerarcnodes[#1](#2)(#3:#4:#5)(#6,#7)
{\coordinate(#6) at ($(#2)+({#5*cos(#3)},{#5*sin(#3)})$);
	\coordinate(#7) at ($(#2)+({#5*cos(#4)},{#5*sin(#4)})$);
	\draw[#1] ($(#2)+({#5*cos(#3)},{#5*sin(#3)})$) arc (#3:#4:#5); }

\def\angcircle(#1)(#2)(#3:#4) {\coordinate(#1) at ($(#2)+({#4*cos(#3)},{#4*sin(#3)})$); }

\tikzset{->-/.style={decoration={
			markings,
			mark=at position #1 with {\arrow{>}}},postaction={decorate}}}

\tikzset{-<-/.style={decoration={
			markings,
			mark=at position #1 with {\arrow{<}}},postaction={decorate}}}

\newcommand{\blink}[2]
{\draw[blue, decoration={markings,mark=at position 0.6 with {\arrow[scale=1.5,>=stealth]{>}}},postaction={decorate}] (#1) --(#2)}

\begin{document}

\title{The Ponzano-Regge cylinder and Propagator for 3d quantum gravity}

\author{{\bf Etera R. Livine}}\email{etera.livine@ens-lyon.fr}
\affiliation{Univ Lyon, Ens de Lyon, Universit\'e Claude Bernard, CNRS,
Laboratoire de Physique, F-69342 Lyon, France}


\date{\today}

\begin{abstract}

We investigate the propagator of 3d quantum gravity, formulated as a discrete topological path integral. We define it as the Ponzano-Regge amplitude of the solid cylinder swept by a  2d disk evolving in time.
Quantum states for a 2d disk live in the tensor products of $N$ spins, where $N$ is the number of holonomy insertions connecting to the disk boundary.
%
We formulate the cylindric amplitude in terms of a transfer matrix and identify its eigen-modes in terms of spin recoupling.
We show that the propagator distinguishes  subspaces with different total recoupled spin. This may select the vanishing overall spin sector at late time depending on the chosen cylinder  boundary data, leading to an emergent symmetry scenario in the continuum limit.
We discuss applications to quantum circuits and the possibility of experimental simulations of this 3d quantum gravity propagator.

\end{abstract}

\maketitle
\tableofcontents

\section*{Introduction}

Three-dimensional gravity can be quantized as a topological quantum field theory (TQFT) \cite{Witten:1988hc}. This is realized in particular by the Ponzano-Regge model \cite{PonzanoRegge1968,Freidel:2004vi,Freidel:2005bb,Barrett:2008wh}. It defines a discretized path integral for the 3d quantum geometry, which can actually be shown to be invariant under the choice of discretization.
It provides us with a local picture of the 3d space-time at the Planck scale, in terms of elementary building blocks, with quantized lengths, areas and volumes. These ``atoms of geometry'' are then glued together to make the 3d space-time, in a diffeomorphism-invariant way. The invariance under diffeomorphism stems from the invariance under changes of discretization\cite{Bonzom:2009zd,Bonzom:2011hm,Bonzom:2011nv,Bonzom:2013tna,Bonzom:2013ofa,Bonzom:2015ans}, constructed out of local moves, e.g. Pachner moves for triangulations, or general bipole and fusion moves for general cellular complexes (e.g. \cite{Girelli:2001wr}).
This formulation proposes an inherently holographic description of 3d gravity with amplitudes depending entirely and exactly only on the boundary state and the bulk topology.

The goal of the present paper is to study the propagator for the basic geometric elements -2d cells-defined by the Ponzano-Regge path integral for 3d quantum gravity. We consider elementary 2d cells, i.e. with the topology of a 2d disk, with a discrete 1d boundary, i.e. triangles (as in the traditional formulation of the Ponzano-Regge model in terms of 3d triangulations) or more general (filled) polygons.
The 2d cell propagates (in time) along a cylinder: the 2d cell sweeps a 3d  solid cylinder, while the 1d circle boundary of the 2d cell (i.e. the ``corner'') sweeps its 2D boundary cylinder.
It turns out that the Ponzano-Regge amplitude for the solid cylinder can be shown to project entirely onto its boundary.

\begin{figure}[h!]

\begin{tikzpicture}[scale=1,every text node part/.style={align=center}]

\coordinate(ai) at (0,0);
\coordinate(af) at (0,4);
\coordinate(bi) at (3,0);
\coordinate(bf) at (3,4);

\coordinate(Ai) at (0,-1.2);
\coordinate(Af) at (0,5);
\coordinate(Bi) at (3,-1.2);
\coordinate(Bf) at (3,5);

\draw[opacity=0] (Ai) to[in=90,out=90,looseness=.5] node[pos=0.3,inner sep=0pt](B1){} node[pos=0.7,inner sep=0pt](B2){}node[pos=0.9,inner sep=0pt](B3){} (Bi) to[,in=-90,out=-90,looseness=.5] node[pos=0.4,inner sep=0pt](B4){} node[pos=0.7,inner sep=0pt](B5){}(Ai);
\draw[opacity=0] (Af) to[in=-90,out=-90,looseness=.5] node[pos=0.1,inner sep=0pt](C1){}node[pos=0.4,inner sep=0pt](C2){} node[pos=0.8,inner sep=0pt](C3){}(Bf) to[in=90,out=90,looseness=.5] node[pos=0.5,inner sep=0pt](C4){} (Af);

\draw[fill=orange, fill opacity=.3] ($(Ai)$) to node[pos=0.5,inner sep=0pt](mb1){} ($(B1)$)to node[pos=0.5,inner sep=0pt](mb2){}($(B2)$)to node[pos=0.5,inner sep=0pt](mb3){}($(B3)$)to node[pos=0.5,inner sep=0pt](mb4){}($(B4)$)to node[pos=0.5,inner sep=0pt](mb5){}($(B5)$)to node[pos=0.5,inner sep=0pt](mb6){}(Ai);
\draw[fill=orange, fill opacity=.3] ($(C1)$) to node[pos=0.5,inner sep=0pt](mc1){}($(C2)$)to node[pos=0.5,inner sep=0pt](mc2){}($(C3)$)to node[pos=0.5,inner sep=0pt](mc3){}($(Bf)$)to node[pos=0.5,inner sep=0pt](mc4){} ($(C4)$)to node[pos=0.5,inner sep=0pt](mc5){} ($(C1)$);

\draw[decoration={markings,mark=at position 1 with {\arrow[scale=1.5,>=stealth]{>}}},postaction={decorate}] ($ (bi)+(1,0)$) -- node[pos=0.5,inner sep=5pt,right]{time} ($ (bf)+(1,0)$);

\draw[blue,thick,decoration={markings,mark=at position 1 with {\arrow[scale=1,>=stealth]{>}}},postaction={decorate}] ($(mb1)+(-0.05,-.2)$) -- ($(mb1)+(-0.05,.2)$) node[ left]{$j_{6}$};
\draw[blue,thick,decoration={markings,mark=at position 1 with {\arrow[scale=1,>=stealth]{>}}},postaction={decorate}] ($(mb2)+(0,-.2)$) -- ($(mb2)+(0,.2)$) node[ left]{$j_{5}$};
\draw[blue,thick,decoration={markings,mark=at position 1 with {\arrow[scale=1,>=stealth]{>}}},postaction={decorate}] ($(mb3)+(0,-.2)$) -- ($(mb3)+(0,.2)$) node[ left]{$j_{4}$};
\draw[blue,thick,decoration={markings,mark=at position 1 with {\arrow[scale=1,>=stealth]{>}}},postaction={decorate}] ($(mb4)+(0,-.2)$) node[ left]{$j_{3}$}-- ($(mb4)+(0,.2)$) ;
\draw[blue,thick,decoration={markings,mark=at position 1 with {\arrow[scale=1,>=stealth]{>}}},postaction={decorate}] ($(mb5)+(0,-.2)$)node[ left]{$j_{2}$} -- ($(mb5)+(0,.2)$) ;
\draw[blue,thick,decoration={markings,mark=at position 1 with {\arrow[scale=1,>=stealth]{>}}},postaction={decorate}] ($(mb6)+(0.05,-.2)$) node[ left]{$j_{1}$}-- ($(mb6)+(0.05,.2)$) ;
\draw[blue,thick,decoration={markings,mark=at position 1 with {\arrow[scale=1,>=stealth]{>}}},postaction={decorate}] ($(mc1)+(0,-.2)$) node[ left]{$j'_{1}$}-- ($(mc1)+(0,.2)$) ;
\draw[blue,thick,decoration={markings,mark=at position 1 with {\arrow[scale=1,>=stealth]{>}}},postaction={decorate}] ($(mc2)+(0,-.2)$) node[ right]{$j'_{2}$}-- ($(mc2)+(0,.2)$) ;
\draw[blue,thick,decoration={markings,mark=at position 1 with {\arrow[scale=1,>=stealth]{>}}},postaction={decorate}] ($(mc3)+(0,-.2)$) node[ right]{$j'_{3}$}-- ($(mc3)+(0,.2)$) ;
\draw[blue,thick,decoration={markings,mark=at position 1 with {\arrow[scale=1,>=stealth]{>}}},postaction={decorate}] ($(mc4)+(0,-.2)$) -- ($(mc4)+(0,.2)$)node[ right]{$j'_{4}$} ;
\draw[blue,thick,decoration={markings,mark=at position 1 with {\arrow[scale=1,>=stealth]{>}}},postaction={decorate}] ($(mc5)+(0,-.2)$) -- ($(mc5)+(0,.2)$)node[ left]{$j'_{5}$} ;

\coordinate(b1) at ($(B1)+(0,1.2)$);
\coordinate(b2) at ($(B2)+(0,1.2)$);
\coordinate(b3) at ($(B3)+(0,1.2)$);
\coordinate(b4) at ($(B4)+(0,1.2)$);
\coordinate(b5) at ($(B5)+(0,1.2)$);
\coordinate(c1) at ($(C1)+(0,-1)$);
\coordinate(c2) at ($(C2)+(0,-1)$);
\coordinate(c3) at ($(C3)+(0,-1)$);
\coordinate(c4) at ($(C4)+(0,-1)$);

\coordinate(x1) at (-.2,1);
\draw (ai)--(x1)--(b5);
\coordinate(x2) at (.3,2.2);
\coordinate(x3) at (-.1,3.4);
\draw(x1)--(x2)--(x3)--(c1);
\coordinate(x4) at (1,1.4);
\draw (x1)--(x4)--(b5);
\coordinate(x5) at (2.3,1.2);
\coordinate(x6) at (3.1,2);
\draw (x4)--(x5)--(b4);
\draw (x5)--(x6)--(b3);
\coordinate(x7) at (.85,2.5);
\draw (x2)--(x7)--(x4);
\coordinate(x8) at (1.95,2.8);
\coordinate(x9) at (2.8,3.2);
\draw(x7)--(x8)--(x5)--(x9)--(x6)--(bf)--(x9)--(c3)--(x8)--(c2)--(x7)--(x3);

\coordinate(y1) at (.7,1.9);
\draw[dotted] (ai)--(y1)--(b1);
\draw[dotted] (x2)--(y1)--(b2)--(x6);
\coordinate(y2) at (1.7,2.3);
\draw[dotted] (y1)--(y2)--(x6);
\coordinate(y3) at (1.9,3.3);
\draw[dotted] (y2)--(y3)--(c4);
\draw[dotted] (y3)--(bf);
\coordinate(y4) at (.45,3.2);
\draw[dotted] (x2)--(y4)--(y3);
\draw[dotted] (y4)--(c1);

\draw[fill=magenta, fill opacity=.3, draw=none] (b3)--(b4)--(b5)--(ai)--(x1)--(x2)--(x3)--(c1)--(c4)--(bf)--(x6)--(b3);

\draw[dotted] (ai)--(b1)--(b2)--(b3);
\draw(b3)--(b4)--(b5)--(ai);
\draw (c1)to (c2) to(c3)to(bf);
\draw[dotted]  (bf) to(c4)to(c1);

\node at ($(Ai)+(-1.5,0)$) {\blue{initial disk}};
\node at ($(Af)+(-1.5,0)$){\blue{final disk}};
\node[text width=2cm] at (-1.5,2) {\color{magenta}{interpolating cylinder}};

\end{tikzpicture}

\caption{The 2d boundary cylinder describing the evolution in time of the boundary state of the initial disk, with spins $j_{1},..,j_{6}$ (encoding the quantized edge lengths in Planck units), to the boundary state of the final disk, with spins $j'_{1},..,j'_{5}$.}
\label{fig:intro}
\end{figure}
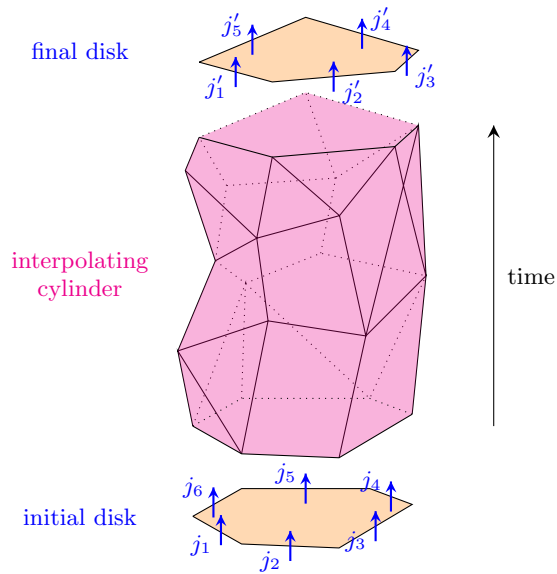

More precisely, the 2d cell's polygonal boundary is made of $N$ edges $e_{i}$, each with a quantized length given by a spin $j_{i}\in\N/2$ in Planck units and carrying a spin state in the corresponding $\SU(2)$ representation $\cV_{j_{i}}$. Then the evolution of those spins is given by a spin network on the 2d boundary cylinder, i.e.  a graph connecting the initial spins to the final spins, dressed spins on the links and $\SU(2)$-invariant tensors (intertwiners) on the nodes, as illustrated on fig.\ref{fig:intro}.
This boundary spin network describes the geometry of the tube or world-sheet swept by the evolving 1d boundary. It corresponds, in a suitable continuum limit, to the 2d boundary metric on the time-like boundary between the initial and final 2d slices.
So the present work explores the computation of such cylinder amplitudes, depending on the 2d boundary data, in the Ponzano-Regge model. We call this the Ponzano-Regge propagator\footnotemark{} for the quantized disk.
\footnotetext{
The Ponzano-Regge propagator, that we define here as the transition amplitude for a 2d disk along a cylinder, clearly depends on the 2d boundary data on the cylinder boundary. 
It is interesting to note the difference with the {\it group field theory} propagator. Indeed, spinfoam amplitudes can be understood as Feynman diagrams of a field theory on a group manifold, or group field theory (GFT) in short \cite{DePietri:1999bx,Reisenberger:2000zc}. The GFT partition function (if renormalizable) then provides a non-perturbative definition of the sum of the spinfoam amplitudes over cellular complexes, thus it defines in particular the sum over bulk topologies. Ponzano-Regge amplitudes are generated by Boulatov's GFT \cite{Boulatov:1992vp} (see also e.g. \cite{Livine:2010zx,Oriti:2014uga} for overviews of the formalism). The GFT propagator, defined as the 2-point function of Boulatov GFT's, also defines a transition amplitude for quantum states of the 2d disk, but it considers discretized 3-manifold whose boundary is strictly the union of two disjoint 2-disks, without any notion of cylindric boundary interpolating between them. The difference between the GFT propagator and the present Ponzano-Regge propagator can thus be summarized as ``no-boundary versus boundary''. It would be interesting to understand if these two notions of propagator for 3d quantum gravity coincide in some regime and limit, e.g. considering tree diagrams for the GFT (as in \cite{Freidel:2005qe}) and considering the large number of time slices for the interpolating cylinder or even the sum over all possible number of time slices (which could shed light on the ongoing discussion of sum over GFT diagrams versus refinement of the spinfoam complexe).
}

The first section sets up the mathematical definition of the Ponzano-Regge propagator and shows that it indeed only depends on the boundary state of the 3d solid cylinder.
The second section frames this propagator in terms of transfer matrix: for a square lattice, the propagator is the composition of transfer matrix associated to each slice and we get the same transfer matrix for every slice upon choosing a homogenous ansatz for the boundary spin network state on the 2d cylinder.
Moreover, we  show that it preserves the total (recoupled) spin of the initial state. This means that a value of the recoupled spin corresponds to a eigenvalue of the propagator, so that the propagator evolve each value  of recoupled spin differently as eigen-modes of the (cylindric) dynamics. For a  large number $S$ of slices, the propagator will then select the largest eigenvalue of the transfer matrix. When this largest eigenvalue is given by a vanishing recoupled spin, this leads to a dynamical selection of $\SU(2)$ invariant states, illustrating a possible mechanism of emergence of symmetry .

The third section is dedicated to working out examples to show how to make the Ponzano-Regge propagator explicit.
We illustrate our general results with the evolution of $N=2$ spins $\f12$ and then of $N=3$ spins $\f12$ interpreted as an elementary quantum triangle. We show in both cases how the propagation of the eigen-modes depend on 2d boundary data given by the boundary spin network. Taking the limit of a large number of slices on a boundary square lattice,  we identify the dominant recoupled spin corresponding to largest eigenvalue of the transfer matrix.

We conclude with a discussion on possible applications of these results beyond the framework of the Ponzano-Regge path integral and 3d quantum gravity, for instance on quantum circuits and possible experimental implementation using condensed matter models. This is especially relevant in the context of potential quantum simulations of loop quantum gravity, intertwiner dynamics and spinfoams, recently discussed in e.g. \cite{Cohen:2020jlj,Czelusta:2020ryq,Zhang:2020lwi,Mielczarek:2021xik}.

\section{The Propagator as a boundary amplitude}

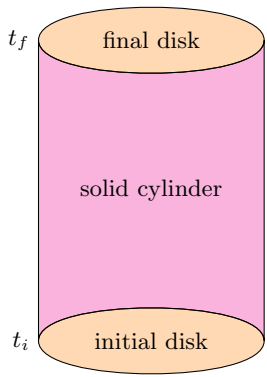
\begin{figure}[h!]
\begin{tikzpicture}[scale=1]

\coordinate(ai) at (0,0);
\coordinate(af) at (0,4);
\coordinate(bi) at (3,0);
\coordinate(bf) at (3,4);


\draw[fill=magenta, fill opacity=.3] (ai) to[dotted,in=90,out=90,looseness=.5] (bi) -- (bf) to[,in=-90,out=-90,looseness=.5] (af) -- (ai);

\draw[fill=orange, fill opacity=.3] (ai) to[dotted,in=90,out=90,looseness=.5] (bi) to[,in=-90,out=-90,looseness=.5] (ai);
\draw[fill=orange, fill opacity=.3] (af) to[dotted,in=90,out=90,looseness=.5] (bf) to[,in=-90,out=-90,looseness=.5] (af);

\node[left] at (ai) {$t_i$};
\node[left] at (af) {$t_f$};
\node at (1.5,2) {solid cylinder};
\node at (1.5,0) {initial disk};
\node at (1.5,4) {final disk};
			
\end{tikzpicture}

\caption{The 3d solid cylinder describing the evolution in time of a 2d disk: it has the topology of a 3-ball; its 2d boundary has the topology of a 2-sphere and can be decomposed into the initial and final disks and a 2d cylinder interpolating between the initial disk boundary and the final disk boundary.}
\label{fig:cylinder}
\end{figure}

The goal of the present work is to study the evolution of quantum states of the 2d disk. Its evolution in time defines a 3d solid cylinder. This region of space-time has the topology of a 3-ball, as depicted on fig.\ref{fig:cylinder}, with a 2-sphere boundary. This boundary sphere should be split into its ``space-like'' boundary, consisting in the initial and final disk, and its ``time-like'' boundary, consisting in the 2d cylinder stretching between the initial disk boundary and the final disk boundary.
Here, we work in a space-time with Riemannian signature (+++), so that space-like or time-like property does not refer to a positive or negative space-time interval but are defined with respect to a coordinate chosen as parameterizing the evolution, thus ``time'', and the corresponding foliation of space-time into 2d spatial slices.

%
If one were to work with a Lorentzian signature (-++), the path integral amplitudes would be given by the Ponzano-Regge spinfoam model with gauge group $\SU(1,1)$ \cite{Freidel:2000uq,Davids:2000kz,Freidel:2002hx,Freidel:2005bb}. The overall geometrical setting of the evolution along the cylinder, from an initial space-like disk to a final space-like disk, would not be modified but the Lorentzian signature would now appear in the difference between the description of the  time-like geometry on the 2d boundary cylinder and the space-like geometry on the canonical initial and final disks. Indeed, the boundary spin network on the time-like boundary would be decorated by $\SU(1,1)$ irreducible representations from the discrete series of unitary representations, while the space-like boundaries would involve the (principal) continuous series of of unitary representations. The Ponzano-Regge amplitudes would then be evaluations of boundary spin networks involving the two series of representations. It would be a significant progress to understand these reflect the Lorentzian signature and causal structure of the quantized space-time (see \cite{Simao:2021qno} for recent work investigating the difference between the Riemannian and Lorentzian spinfoam amplitudes in 4d quantum gravity). We postpone such study to future investigation and focus here on the Riemann signature model with gauge group $\SU(2)$.

%
%
%
%
%

\subsection{Boundary states}

\begin{figure}[h!]

\begin{tikzpicture}[scale=1.2]

\coordinate(a) at (0,0) ;

\draw (a) node {$\bullet$} ++(0,0.3) node{$I_{v}$};

\draw[decoration={markings,mark=at position 0.7 with {\arrow[scale=1.5,>=stealth]{<}}},postaction={decorate}]  (a)--++(-.85,0.3) node[left]{$j_{2}$};
\draw[decoration={markings,mark=at position 0.7 with {\arrow[scale=1.5,>=stealth]{<}}},postaction={decorate}]  (a)--++(-.8,0.7) node[left]{$j_{1}$};
\draw[decoration={markings,mark=at position 0.7 with {\arrow[scale=1.5,>=stealth]{<}}},postaction={decorate}]  (a)--++(-.8,-.7)  ;

\draw[decoration={markings,mark=at position 0.7 with {\arrow[scale=1.5,>=stealth]{>}}},postaction={decorate}]  (a)--++(.85,0.3) node[right]{$j'_{2}$};
\draw[decoration={markings,mark=at position 0.7 with {\arrow[scale=1.5,>=stealth]{>}}},postaction={decorate}]  (a)--++(.8,0.7) node[right]{$j'_{1}$};
\draw[decoration={markings,mark=at position 0.7 with {\arrow[scale=1.5,>=stealth]{>}}},postaction={decorate}]  (a)--++(.8,-.7)  ;

\draw[dotted,line width=1pt] (-.95,0) -- (-.95,-.4);
\draw[dotted,line width=1pt] (.95,0) -- (.95,-.4);

\node at (-2.5,0) {$\displaystyle{\bigotimes_{e|v=t(e)}\cV_{j_{e}}}$};
\node at (+2.5,0) {$\displaystyle{\bigotimes_{e|v=s(e)}\cV_{j_{e}}}$};
\end{tikzpicture}

\caption{An intertwiner living on a spin network vertex $v$ as a $\SU(2)$-invariant map between the incoming spins and the outgong spins.}
\label{fig:intertwiner}
\end{figure}
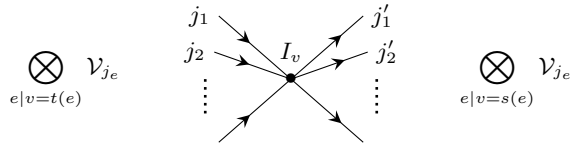

The quantum geometry on the 2-sphere, thought as the boundary of the 3-ball, is defined as a spin network state, which can be understood as a quantized discrete geometry. A spin network state is defined on an oriented graph on the 2-sphere. On that graph $\Gamma$, we consider the holonomies of the (pull-back of the) 3d spin connection, which describes the transport of vectors in space-time, along the graph edges. Those holonomies are $\SU(2)$ group elements, $g_{e}\in \SU(2)$ on the oriented edge $e$. A spin network state on the graph $\Gamma$ is defined as a wave-function of those holonomies:
\be
\psi(\{g_{e}\}_{e\in\Gamma})\quad\in\quad
\cF(\SU(2)^{E})\,,
\ee
where we call $E$ the number of edges of the graph $\Gamma$. We further require these wave-functions to be invariant under $\SU(2)$ transformations at every vertex of the graph, which implements the gauge-invariance of the state under local $\SU(2)$  transformations:
\be
\psi(\{g_{e}\}_{e\in\Gamma})=\psi(\{h_{t(e)}g_{e}h_{s(e)}^{-1}\}_{e\in\Gamma})\,,\qquad
\forall \{h_{v}\}_{v\in\Gamma}\in\SU(2)^{V}\,,
\ee
where $V$ is the number of vertices of the graph $\Gamma$, and $s(e)$, respectively $t(e)$, is the source, respectively target, of the oriented edge $e$. And we endow these wave-functions with a scalar product:
\be
\la \psi|\tpsi\ra_{\Gamma}=\int_{\SU(2)^{E}} \prod_{e}\rd g_{e}\,\overline{\psi(\{g_{e}\})}\tpsi(\{g_{e}\})\,,
\ee
where we use the Haar measure on $\SU(2)$.
This defines the Hilbert space of spin network states living on a graph $\Gamma$ on the boundary of a 3-ball:
\be
\cH_{\Gamma}=L^{2}(\SU(2)^{E}/\SU(2)^{V})\,.
\ee
Spin network basis states are provided by decomposing $L^{2}$ functions on $\SU(2)$ into Wigner matrices. This decomposition amounts to a Fourier transform on the $\SU(2)$ Lie group:
\be
f\in L^{2}(\SU(2))
\quad\Rightarrow\quad
f(g)=\sum_{j,m,n} f^{(j)}_{mn}D^{j}_{mn}(g)
\quad\textrm{with}\,\,
\int_{\SU(2)} \rd g\,|f(g)|^{2}=\sum_{j}\f1{2j+1}\sum_{m,n} | f^{(j)}_{mn}|^{2}
<+\infty
\,.
\ee
The half-integer $j\in\N/2$ is called the spin and labels the unitary irreducible representations of $\SU(2)$. The representation of spin $j$ defines a Hilbert space $\cV_{j}$ of dimension $\dim \cV_{j}=2j+1$, with its usual basis states $|j,m\ra$ labeled by the spin $j$ and the magnetic moment $m$ running from $-j$ to $+j$ by integer step. These basis states diagonalize both the $\su(2)$ Casimir ($\vJ^{2}$) and the generator of $\u(1)$ rotations in a chosen direction (usually $J_{3}$). Group elements are represented as $(2j+1)\times(2j+1)$ matrices acting in the $|j,m\ra$ basis. These are the Wigner matrices:
\be
D^{j}_{mn}(g)=\la j,m|g|j,n\ra\,.
\ee
These Fourier modes on $\SU(2)$ allow to define the spin network basis states: 
\be
\Phi^{\{j_{e},I_{v}\}}(\{g_{e}\})
\equiv
\prod_{e}D^{j_{e}}_{m^{t}_{e}m^{s}_{e}}(g_{e})
\,
\prod_{v}
\la \bigotimes_{e|v=s(e)}j_{e},m^{s}_{e}\,|I_{v}|\,\bigotimes_{e|v=t(e)}j_{e},m^{t}_{e}\ra
\,,
\ee
where a spin network basis state is defined by the assignment of spins $j_{e}$ to every edge $e$ and of intertwiners $I_{v}$ to every vertex.
As illustrated on fig.\ref{fig:intertwiner}, an intertwiner is a $\SU(2)$-invariant map, i.e. here:
\be
I_{v}:\bigotimes_{e|v=t(e)}\cV_{j_{e}}\rightarrow \bigotimes_{e|v=s(e)}\cV_{j_{e}}
\qquad\textrm{such that}\quad
g\circ I_{v}=I_{v}\circ g\,,\,\,\forall g\in\SU(2)\,.
\ee
In short, an intertwiner is a singlet state living in the tensor product of the incoming and outgoing spins at the vertex $v$:
\be
I_{v}\in \textrm{Inv}_{\SU(2)}\Big{[}\bigotimes_{e|v=t(e)}\cV_{j_{e}}\otimes \bigotimes_{e|v=s(e)}\cV_{j_{e}}^{*}\Big{]}
\,.
\ee

\begin{figure}[h!]

\begin{subfigure}[t]{0.45\linewidth}
\begin{tikzpicture}[scale=3.5]

\coordinate (A1) at (0,0);
\coordinate (A2) at (0.5,-0.1);
\coordinate (A3) at (.65,.4);
\coordinate (A4) at (.3,.4);

\draw (A1)--  node[pos=0.5,inner sep=0pt](m12){} (A2);
\draw (A2)--  node[pos=0.5,inner sep=0pt](m23){}(A3) --node[pos=0.5,inner sep=0pt](m34){} (A4) -- node[pos=0.5,inner sep=0pt](m14){} (A1);
\fill[fill=black,fill opacity=0.05]  (A1) -- (A2) -- (A3) -- (A4);

\coordinate (A5) at (.95,0);
\draw (A2) --  node[pos=0.5,inner sep=0pt](m25){} (A5) --  node[pos=0.5,inner sep=0pt](m35){} (A3);
\fill[fill=black,fill opacity=0.05]  (A2) -- (A5) -- (A3);

\coordinate (A6) at (1.1,.65);
\draw (A5) --  node[pos=0.5,inner sep=0pt](m56){} (A6) --  node[pos=0.5,inner sep=0pt](m36){} (A3);
\fill[fill=black,fill opacity=0.05]  (A5) -- (A6) -- (A3);

\coordinate (A7) at (.57,.8);
\draw (A3) --  node[pos=0.5,inner sep=0pt](m37){} (A7) -- node[pos=0.5,inner sep=0pt](m67){}(A6);
\fill[fill=black,fill opacity=0.05]  (A3) -- (A7) -- (A6);
\draw (A4) -- node[pos=0.5,inner sep=0pt](m47){}(A7);
\fill[fill=black,fill opacity=0.05]  (A4) -- (A7) -- (A3);

\coordinate (A8) at (.93,1.05);
\draw (A6) --node[pos=0.5,inner sep=0pt](m68){} (A8) -- node[pos=0.5,inner sep=0pt](m78){}(A7);
\fill[fill=black,fill opacity=0.05]  (A6) -- (A8) -- (A7);

\coordinate (A9) at (.62,1.2);
\draw (A8) -- node[pos=0.5,inner sep=0pt](m89){}(A9) -- node[pos=0.5,inner sep=0pt](m79){}(A7);
\fill[fill=black,fill opacity=0.05] (A8) -- (A9) -- (A7);

\coordinate (A10) at (-0.1,.75);
\draw (A9) -- node[pos=0.5,inner sep=0pt](m910){}(A10) -- node[pos=0.5,inner sep=0pt](m710){}(A7);
\fill[fill=black,fill opacity=0.05] (A9) -- (A10) -- (A7);
\draw (A10) -- node[pos=0.5,inner sep=0pt](m410){}(A4);
\fill[fill=black,fill opacity=0.05] (A4) -- (A10) -- (A7);
\draw (A10) -- node[pos=0.5,inner sep=0pt](m110){}(A1);
\fill[fill=black,fill opacity=0.05] (A4) -- (A10) -- (A1);

\coordinate (a1) at (.73,.65);
\draw[dotted] (A8) -- node[pos=0.5,inner sep=0pt](N18){}(a1) -- node[pos=0.5,inner sep=0pt](N15){}(A5);
\coordinate (a2) at (.33,.6);
\draw[dotted] (A9) -- node[pos=0.5,inner sep=0pt](N19){}(a1) -- node[pos=0.5,inner sep=0pt](n12){}(a2) -- node[pos=0.5,inner sep=0pt](N210){}(A10);
\coordinate (a3) at (.55,.3);
\draw[dotted] (a1) --node[pos=0.5,inner sep=0pt](n13){} (a3) -- node[pos=0.5,inner sep=0pt](n23){}(a2);
\draw[dotted] (A1) -- node[pos=0.5,inner sep=0pt](N13){}(a3) --node[pos=0.5,inner sep=0pt](N35){} (A5) ;

\coordinate (B1) at (.4,.15);
\draw (B1) node{\blue\tiny{$\bullet$}};
\coordinate (B2) at (.9,.35);
\draw (B2) node{\blue\tiny{$\bullet$}};
\coordinate (B3) at (.1,.35);
\draw (B3) node{\blue\tiny{$\bullet$}};
\coordinate (B4) at (.33,.92);
\draw (B4) node{\blue\tiny{$\bullet$}};
\coordinate (b1) at (.75,1);
\draw[opacity=.2] (b1) node{\blue\tiny{$\bullet$}};
\coordinate (b123) at ($.33*(a1)+.33*(a2)+.33*(a3)$);
\draw[opacity=.2] (b123) node{\blue\tiny{$\bullet$}};
\coordinate (b1235) at ($.25*(a3)+.25*(A1)+.25*(A5)+.25*(A2)$);
\draw[opacity=.2] (b1235) node{\blue\tiny{$\bullet$}};
\coordinate (b12) at ($.25*(a1)+.25*(a2)+.25*(A9)+.25*(A10)$);
\draw[opacity=.2] (b12) node{\blue\tiny{$\bullet$}};
\coordinate (b158) at ($.25*(a1)+.25*(A5)+.25*(A8)+.25*(A6)$);
\draw[opacity=.2] (b158) node{\blue\tiny{$\bullet$}};
\coordinate (b135) at ($.33*(a1)+.33*(A5)+.33*(a3)$);
\draw[opacity=.2] (b135) node{\blue\tiny{$\bullet$}};
\coordinate (b23) at ($.25*(a3)+.25*(A1)+.25*(a2)+.25*(A10)$);
\draw[opacity=.2] (b23) node{\blue\tiny{$\bullet$}};

\coordinate (B5) at ($.33*(A4)+.33*(A7)+.33*(A10)$);
\draw (B5) node{\blue\tiny{$\bullet$}};
\coordinate (B6) at ($.33*(A4)+.33*(A7)+.33*(A3)$);
\draw (B6) node{\blue\tiny{$\bullet$}};
\coordinate (B7) at ($.33*(A6)+.33*(A7)+.33*(A3)$);
\draw (B7) node{\blue\tiny{$\bullet$}};
\coordinate (B8) at ($.33*(A2)+.33*(A5)+.33*(A3)$);
\draw (B8) node{\blue\tiny{$\bullet$}};
\coordinate (B9) at ($.33*(A6)+.33*(A7)+.33*(A8)$);
\draw (B9) node{\blue\tiny{$\bullet$}};
\coordinate (B10) at ($.33*(A9)+.33*(A7)+.33*(A8)$);
\draw (B10) node{\blue\tiny{$\bullet$}};

\draw[blue] (B1)--(m12);
\draw[blue] (B1)--(m23);
\draw[blue] (B1)--(m14);
\draw[blue] (B1)--(m34);

\draw[blue] (B2)--(m35);
\draw[blue] (B2)--(m56);
\draw[blue] (B2)--(m36);

\draw[blue] (B3)--(m14);
\draw[blue] (B3)--(m110);
\draw[blue] (B3)--(m410);

\draw[blue] (B4)--(m79);
\draw[blue] (B4)--(m710);
\draw[blue] (B4)--(m910);

\draw[blue] (B5)--(m47);
\draw[blue] (B5)--(m410);
\draw[blue] (B5)--(m710);

\draw[blue] (B6)--(m47);
\draw[blue] (B6)--(m34);
\draw[blue] (B6)--(m37);

\draw[blue] (B7)--(m67);
\draw[blue] (B7)--(m36);
\draw[blue] (B7)--(m37);

\draw[blue] (B8)--(m23);
\draw[blue] (B8)--(m25);
\draw[blue] (B8)--(m35);

\draw[blue] (B9)--(m67);
\draw[blue] (B9)--(m78);
\draw[blue] (B9)--(m68);

\draw[blue] (B10)--(m78);
\draw[blue] (B10)--(m79);
\draw[blue] (B10)--(m89);

\draw[blue, opacity=.2] (b123)--(n12);
\draw[blue, opacity=.2] (b123)--(n23);
\draw[blue, opacity=.2] (b123)--(n13);

\draw[blue, opacity=.2] (b1235)--(m12);
\draw[blue, opacity=.2] (b1235)--(m25);
\draw[blue, opacity=.2] (b1235)--(N35);
\draw[blue, opacity=.2] (b1235)--(N13);

\draw[blue, opacity=.2] (b135)--(N35);
\draw[blue, opacity=.2] (b135)--(n13);
\draw[blue, opacity=.2] (b135)--(N15);

\draw[blue, opacity=.2] (b158)--(N15);
\draw[blue, opacity=.2] (b158)--(N18);
\draw[blue, opacity=.2] (b158)--(m56);
\draw[blue, opacity=.2] (b158)--(m68);

\draw[blue, opacity=.2] (b1)--(m89);
\draw[blue, opacity=.2] (b1)--(N19);
\draw[blue, opacity=.2] (b1)--(N18);

\draw[blue, opacity=.2] (b12) --(n12);
\draw[blue, opacity=.2] (b12) --(N210);
\draw[blue, opacity=.2] (b12) --(N19);
\draw[blue, opacity=.2] (b12) --(m910);

\draw[blue, opacity=.2] (b23) --(m110);
\draw[blue, opacity=.2] (b23) --(n23);
\draw[blue, opacity=.2] (b23) --(N13);
\draw[blue, opacity=.2] (b23) --(N210);

\end{tikzpicture}
\caption{Cellular decomposition of the 2-sphere (in black) and its dual graph (in {\blue blue}) on which live the spin network states: each face is represented to a spin network vertex ${\scriptstyle\blue \bullet}$ decorated by a $\SU(2)$ intertwiner.}

\end{subfigure}
\hspace*{8mm}
\begin{subfigure}[t]{0.45\linewidth}

	\begin{tikzpicture}[scale=2]

		\coordinate (A1) at (-0.39,0.81);
		\coordinate (A2) at (-0.55,0.19);
		\coordinate (A3) at (-0.15,-0.58);
		\coordinate (A4) at (0.4,-0.78);
		\coordinate (A5) at (1.21,-0.58);
		\coordinate (A6) at (2.04,-0.83);
		\coordinate (A7) at (1.79,0.11);
		\coordinate (A8) at (1.83, 0.74);
		\coordinate (A9) at (1.26, 0.76);
		\coordinate (A10) at (0.64, 1.01);
		\coordinate (I1) at (0.28, -0.14);
		\coordinate (I2) at (0.67, 0.24);
		\coordinate (I3) at (1.34, 0.06);
		\coordinate (I4) at (1.12, -0.22);
		
		
		\draw(A1)--(A2);
		\draw(A2)--(A3);
		\draw(A3)--(A4);
		\draw(A4)--(A5);
		\draw(A5)--(A6);
		\draw(A6)--(A7);
		\draw(A7)--(A8);
		\draw(A8)--(A9);
		\draw(A9)--(A10);
		\draw(A10)--(A1);
		\draw(A1)--(I1);
		\draw(A2)--(I1);
		\draw(A3)--(I1);
		\draw(A4)--(I1); 
		\draw(I1)--(I2);
		\draw(I2)--(A10);
		\draw(I2)--(I3);
		\draw(I3)--(A9);
		\draw(I3)--(A7);
		\draw(I3)--(I4);
		\draw(I4)--(A5);
		\draw(I4)--(I2);    
		
		\coordinate (B1) at (0.27, 0.52); \coordinate (B2) at (0.98, 0.49); \coordinate (B3) at (1.56, 0.46); \coordinate (B4) at (1.6, -0.35); \coordinate (B5) at (1.06, 0.02); \coordinate (B6) at (0.75, -0.34); \coordinate (B7) at (0.2, -0.48); \coordinate (B8) at (-0.2, -0.2); \coordinate (B9) at (-0.2, 0.24); \coordinate (B10) at (0.03, 1.21); \coordinate (B11) at (1.03, 1.19); \coordinate (B12) at (1.57, 1.1); \coordinate (B13) at (2.06, 0.44); \coordinate (B14) at (2.16, -0.27); \coordinate (B15) at (1.6, -1); \coordinate (B16) at (0.92, -1.0); \coordinate (B17) at (-0.06, -1.02); \coordinate (B18) at (-0.62, -0.44); \coordinate (B19) at (-0.71, 0.59); 
		
\blink{B1}{B2};
\blink{B1}{B9};
\blink{B10}{B1};
\blink{B1}{B6};
\blink{B3}{B2};
\blink{B3}{B13};
\blink{B3}{B12};
\blink{B2}{B5};
\blink{B11}{B2};
\blink{B3}{B4};
\blink{B4}{B5};
\blink{B4}{B6};
\blink{B14}{B4};
\blink{B4}{B15};
\blink{B5}{B6};
\blink{B7}{B6};
\blink{B6}{B16};
\blink{B7}{B8};
\blink{B7}{B17};
\blink{B8}{B9};
\blink{B9}{B19};
\blink{B18}{B8};

\node[below left] at (B10) {{\blue $j_{e}$}}; 
		
		\draw[blue] (B1) node{$\bullet$} node[above right]{$I_{v}$}; \draw[blue] (B2) node{$\bullet$}; 
		\draw[blue] (B3) node{$\bullet$}; \draw[blue] (B4) node{$\bullet$}; 
		\draw[blue] (B5) node{$\bullet$}; \draw[blue] (B6) node{$\bullet$}; 
		\draw[blue] (B7) node{$\bullet$}; \draw[blue] (B8) node{$\bullet$}; 
		\draw[blue] (B9) node{$\bullet$};
	\end{tikzpicture}

\caption{A flatten portion of the cellular decomposition of the 2-sphere and the spin network state living on the dual graph dressed with spins $j_{e}\in\f\N2$ along the links and intertwiners $I_{v}$ at the vertices: a spin gives the quantized length in Planck units of the dual edge of the cellular complexe while the intertwiner defines a quantized polygon.}

\end{subfigure}

\caption{A spin network state on the 2-sphere (in {\blue blue}) and the corresponding dual discrete geometry (in black).}
\label{fig:spinnetwork}
\end{figure}
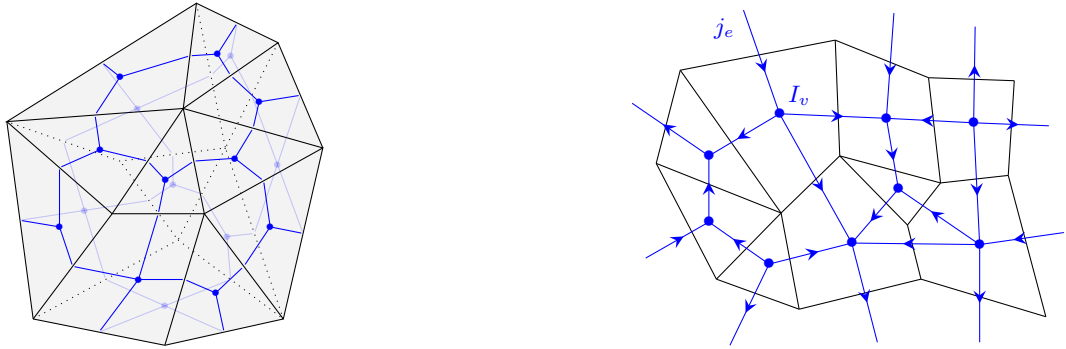

Spin network basis states have a natural interpretation as quantized discrete geometries. On the 2-sphere, one draws the dual cellular complex, as illustrated on fig.\ref{fig:spinnetwork}: each vertex defines a dual face, each edge defines a dual edge between two dual faces, each elementary loop of the graph defines a dual point. Then the spins $j_{e}$ are interpreted as the edge lengths in Planck units on this cellular complex, while the intertwiners $I_{v}$ are understood as quantized polygons and carry the  information about the shape and area of the faces of the  cellular complex (see on the geometrical interpretation of spin networks in 3d quantum gravity e.g. \cite{Rovelli:1993kc,Freidel:2002hx,Bonzom:2011hm,Livine:2013tsa}).

The simplest configurations consist in 3-valent graphs: every 3-valent vertex is a dual to a triangle, every 3-valent intertwiner is uniquely determined by its 3 spins and defines a quantum triangle (see e.g. \cite{Baez:1999tk}) and, thus,  every 3-valent spin network basis state is interpreted as a quantized triangulation of the 2-sphere. Extending this picture, every 4-valent intertwiner defines a quantum quadrilateral, and so on as we add extra links attached to the vertex.

\medskip

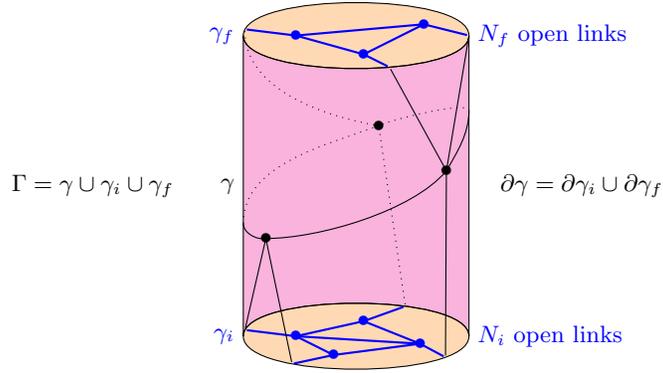
\begin{figure}[h!]

\begin{tikzpicture}[scale=1]

\coordinate(ai) at (0,0);
\coordinate(af) at (0,4);
\coordinate(bi) at (3,0);
\coordinate(bf) at (3,4);

\draw[fill=magenta, fill opacity=.3] (ai) to[dotted,in=90,out=90,looseness=.5] (bi) -- (bf) to[,in=-90,out=-90,looseness=.5] (af) -- (ai);
\draw[fill=orange, fill opacity=.3] (ai) to[dotted,in=90,out=90,looseness=.5] node[pos=0.05,inner sep=0pt](b1){} node[pos=0.65,inner sep=0pt](b2){} (bi) to[,in=-90,out=-90,looseness=.5] node[pos=0.2,inner sep=0pt](b4){} node[pos=0.7,inner sep=0pt](b3){}(ai);
\draw[fill=orange, fill opacity=.3] (af) to[dotted,in=90,out=90,looseness=.5] node[pos=0.05,inner sep=0pt](c1){} (bf) to[,in=-90,out=-90,looseness=.5] node[pos=0,inner sep=0pt](c2){} node[pos=0.4,inner sep=0pt](c3){} (af);


\coordinate(v1) at (0.7,0);
\coordinate(v2) at (1.6,0.2);
\coordinate(v3) at (1.2,-0.25);
\coordinate(v4) at (2.35,-0.1);

\draw (v1) node {\blue{$\bullet$}};
\draw (v2) node {\blue{$\bullet$}};
\draw (v3) node {\blue{$\bullet$}};
\draw (v4) node {\blue{$\bullet$}};
\draw[thick,blue] (v1)--(v2)--(v4)--(v3)--(v1);
\draw[thick,blue] (v1)--(v4);

\coordinate(w1) at (0.7,4+0);
\coordinate(w2) at (2.4, 4+0.15);
\coordinate(w3) at (1.6,4+-0.25);
\draw (w1) node {\blue{$\bullet$}};
\draw (w2) node {\blue{$\bullet$}};
\draw (w3) node {\blue{$\bullet$}};
\draw[thick,blue] (w1)--(w2)--(w3)--(w1);

\draw[thick,blue] (v1)--(b1);
\draw[thick,blue] (v2)--(b2);
\draw[thick,blue] (v3)--(b3);
\draw[thick,blue] (v4)--(b4);

\draw[thick,blue] (w1)--(c1);
\draw[thick,blue] (w2)--(c2);
\draw[thick,blue] (w3)--(c3);

\coordinate(x1) at (0.3,1.3);
\coordinate(x2) at (1.8,2.8);
\coordinate(x3) at (2.7,2.2);
\draw (x1) node {{$\bullet$}};
\draw (x2) node {{$\bullet$}};
\draw (x3) node {{$\bullet$}};
\draw(x1) to[in=-130,out=0,looseness=.7]  (x3);

\draw(c3)--(x3)--(b4);
\draw(x3)--(c2);
\draw[dotted] (b2)--(x2) ;
\draw(b3)--(x1)--(b1);
\draw[dotted] (c1) to[in=170,out=-90,looseness=.9]  (x2);
\draw (x1) to[in=-90,out=-170,looseness=.9] (0,1.5) ;
\draw[dotted] (0,1.5) to[in=-160,out=90,looseness=.9]  (x2);
\draw (x3) to[in=-90,out=60,looseness=.9] (3,3) ;
\draw[dotted] (3,3) to[in=30,out=90,looseness=.3]  (x2);

\node[left] at (ai) {\blue{$\gamma_i$}};
\node[left] at (af) {\blue{$\gamma_f$}};
\node[left] at (0,2) {{$\gamma$}};

\node at (-2,2) {{$\Gamma=\gamma\cup\gamma_{i}\cup\gamma_{f}$}};
\node at (4.5,2) {{$\pp\gamma=\pp\gamma_{i}\cup\pp\gamma_{f}$}};

\node[right] at (bi) {\blue{$N_i$ open links}};
\node[right] at (bf) {\blue{$N_f$ open links}};

\end{tikzpicture}

\caption{Splitting of the boundary spin network on the 2-sphere into the  open spin networks on the initial and final disks and the spin network on the 2d cylinder.}
\label{fig:graphdecomposition}
\end{figure}

Now that we have spin network states on the solid cylinder boundary as a 2-sphere, we would like to decompose this boundary state into canonical boundary and evolving boundary. The canonical boundary consists in the initial and final disks, while the evolving boundary is the cylinder describing the evolution in time of the disk's boundary.
Thus considering a spin network state on the 2-sphere and its underlying (planar) graph $\Gamma$, we cut out the initial and final disks: we get the graph $\gamma_{i}$ for the initial disk, the graph $\gamma_{f}$ for the final disk and the graph $\gamma$ for the 2d cylinder viewed as a sphere with two punctures, as illustrated on fig.\ref{fig:graphdecomposition}. The initial disk graph $\gamma_{i}$ has $N_{i}$ open links, the final disk graph $\gamma_{f}$ has $N_{f}$ open links, and the cylinder graph $\gamma$ has $N_{i}+N_{f}$ open links.

In terms of $\SU(2)$ holonomies, distinguishing edges in the initial graph's interior $\gamma^o_{i}$ and edges on the initial graph's boundary $\pp\gamma_{i}$, the initial spin network state on $\gamma_{i}$ is a wave-function:
\be
\label{psii}
\psi_{i}\big{(}{\{g_{e}\}_{e\in\gamma^o_{i}}, \{H_{e}\}_{e\in\pp\gamma_{i}}}\big{)}
=
\psi_{i}\big{(}{\{h_{t(e)}g_{e}h_{s(e)}^{-1}\}_{e\in\gamma^o_{i}}, \{H_{e}h_{s(e)}^{-1}\}_{e\in\pp\gamma_{i}}}\big{)}
\,,\qquad \forall h_{v}\in\gamma_{i}\,,
\ee
where we have assumed in order to keep simple conventions that all the boundary edges are outgoing.
Similarly, we require the final spin network state to be gauge-invariant under $\SU(2)$ transformations at every vertex of the final disk graph $\gamma_{f}$:
\be
\psi_{f}\big{(}{\{g_{e}\}_{e\in\gamma^o_{f}}, \{H_{e}\}_{e\in\pp\gamma_{f}}}\big{)}
=
\psi_{i}\big{(}{\{h_{t(e)}g_{e}h_{s(e)}^{-1}\}_{e\in\gamma^o_{f}}, \{h_{t(e)}H_{e}\}_{e\in\pp\gamma_{f}}}\big{)}
\,,\qquad \forall h_{v}\in\gamma_{f}\,,
\ee
where we have assumed in order to keep simple conventions that all the boundary edges are incoming.
Then the boundary spin network state on the 2d cylinder is similarly a wave-function:
\be
\label{psiboundary}
\Psi\big{(}{\{g_{e}\}_{e\in\gamma^o}, \{H_{e}\}_{e\in\pp\gamma}}\big{)}
=
\Psi\big{(}{\{h_{t(e)}g_{e}h_{s(e)}^{-1}\}_{e\in\gamma^o}, \{h_{t(e)}H_{e}\}_{e\in\pp\gamma_{i}}, \{H_{e}h_{s(e)}^{-1}\}_{e\in\pp\gamma_{f}}}\big{)}\,,
\ee
where the cylinder boundary $\pp\gamma$ is to be identified with the initial and final graph boundaries $\pp\gamma_{i}\cup \pp\gamma_{f}$. This identification is done through boundary data defined as $\SU(2)$ holonomies along the corner links  connecting the disk graphs to the cylinder graph, $e\in\pp\gamma_{i}\cup \pp\gamma_{f} \sim \pp\gamma$. This allows to define the spin network wave-function $\psi$ on the 2-sphere as a convolution product of the canonical spin networks $\psi_{i,f}$ and the cylinder spin network:
\beq
\label{glueg}
\psi(\{g_{e}\}_{e\in\Gamma})
&=&
\psi(\{g_{e}\}_{e\in\gamma\cup\gamma_{i}\cup\gamma_{f}},\{g_{e}\}_{e\in\pp\gamma})\\
&=&
\int \prod_{e\in\pp\gamma}\rd H_{e}\,
\psi_{i}\big{(}\{g_{e}\}_{e\in\gamma^o_{i}}, \{H_{e}\}_{e\in\pp\gamma_{i}}\big{)}
\psi_{f}\big{(}\{g_{e}\}_{e\in\gamma^o_{f}}, \{H_{e}^{-1}\}_{e\in\pp\gamma_{f}}\big{)}
\,
\Psi\big{(}\{g_{e}\}_{e\in\gamma^o}, \{g_{e}H_{e}^{-1}\}_{e\in\pp\gamma_{i}},\{H_{e}g_{e}\}_{e\in\pp\gamma_{f}}\big{)}
\,,\nn
\eeq
where the boundary data $\{g_{e}\}_{e\in\pp\gamma}$ allows to glue the disk states with the cylinder state into an overall boundary state on the 2-sphere. This boundary  data lives on the boundary of the disks, i.e. on the corners.

\medskip

In order to work with clear geometrical interpretation of the boundary geometry and compute the corresponding Ponzano-Regge amplitudes, it is convenient to switch to the spin basis. We adapt the definition of spin network basis states to graphs with open ends for wave-functions satisfying (\ref{psii}-\ref{psiboundary}), i.e. invariant under gauge transformations of bulk edges and covariant under gauge transformations of boundary edges.
Considering the initial disk, a spin network basis state on the graph $\gamma_{i}$ with boundary is labeled by spins $\{j_{e}\}$ on every edge $e\in\gamma_{i}$ and consists in  the tensor product of intertwiner states $I_{v}$ at every vertex $v\in\gamma_{i}$ with spin states $|j_{e},m_{e}\ra$ for every boundary edge $e\in\pp\gamma_{i}$:
\beq
\label{gluespin}
\Phi^{\{j_{e\in\gamma_{i}},I_{v},m_{e\in\pp\gamma_{i}}\}}_{\gamma_{i}}
(\{g_{e}\}_{e\in\gamma_{i}^o},\{H_{e}\}_{e\in\pp\gamma_{i}})
&\equiv
\displaystyle{\sum_{\{m_{e}^{s,t}\}_{e\in\gamma_{i}^o}}\sum_{\{m_{e}^{s}\}_{e\in\pp\gamma_{i}}}}
&
\prod_{e\in\gamma_{i}^o}D^{j_{e}}_{m^{t}_{e}m^{s}_{e}}(g_{e})
\prod_{e\in\pp\gamma_{i}}D^{j_{e}}_{m_{e}m^{s}_{e}}(g_{e})
\nn\\
&&\prod_{v}
\la \bigotimes_{e|v=s(e)}j_{e},m^{s}_{e}\,|I_{v}|\,\bigotimes_{e|v=t(e)}j_{e},m^{t}_{e}\ra
\,.
\eeq
As formalized in \cite{Chen:2021vrc}, this wave-function $\Phi^{\{j_{e\in\gamma},I_{v},m_{e\in\pp\gamma}\}}_{\gamma}$ can also be understood as a function $\Phi^{\{j_{e\in\gamma},I_{v}\}}_{\gamma}$ mapping the bulk holonomies $\{g_{e}\}_{e\in\gamma_{i}^o}$ to boundary states in the boundary Hilbert space $\otimes_{e\in\pp\gamma}\cV_{j_{e}}$.
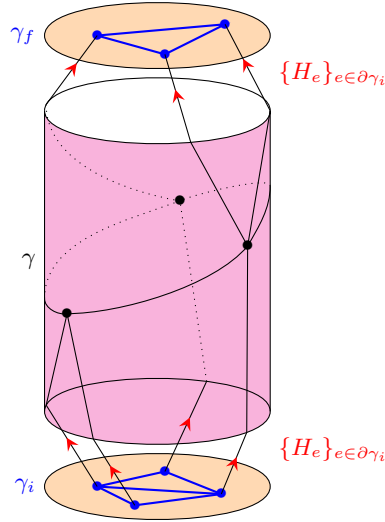
\begin{figure}[h!]

\begin{tikzpicture}[scale=1]

\coordinate(ai) at (0,0);
\coordinate(af) at (0,4);
\coordinate(bi) at (3,0);
\coordinate(bf) at (3,4);

\coordinate(Ai) at (0,-1);
\coordinate(Af) at (0,5);
\coordinate(Bi) at (3,-1);
\coordinate(Bf) at (3,5);

\draw[fill=magenta, fill opacity=.3] (ai) to[dotted,in=-90,out=-90,looseness=.5] (bi) -- (bf) to[,in=-90,out=-90,looseness=.5] (af) -- (ai);
\draw (ai) to[in=90,out=90,looseness=.5] (bi) ;
\draw (af) to[in=90,out=90,looseness=.5] (bf);

\draw[fill=orange, fill opacity=.3] (Ai) to[dotted,in=90,out=90,looseness=.5] node[pos=0.05,inner sep=0pt](B1){} node[pos=0.65,inner sep=0pt](B2){} (Bi) to[,in=-90,out=-90,looseness=.5] node[pos=0.2,inner sep=0pt](B4){} node[pos=0.7,inner sep=0pt](B3){}(Ai);
\draw[fill=orange, fill opacity=.3] (Af) to[dotted,in=90,out=90,looseness=.5] node[pos=0.05,inner sep=0pt](C1){} (Bf) to[,in=-90,out=-90,looseness=.5] node[pos=0,inner sep=0pt](C2){} node[pos=0.4,inner sep=0pt](C3){} (Af);

\coordinate(bf) at ($(Bf)+(0,-1)$);
\coordinate(bi) at ($(Bi)+(0,1)$);
\coordinate(af) at ($(Af)+(0,-1)$);
\coordinate(ai) at ($(ai)+(0,1)$);

\coordinate(b1) at ($(B1)+(0,1)$);
\coordinate(b2) at ($(B2)+(0,1)$);
\coordinate(b3) at ($(B3)+(0,1)$);
\coordinate(b4) at ($(B4)+(0,1)$);
\coordinate(c1) at ($(C1)+(0,-1)$);
\coordinate(c2) at ($(C2)+(0,-1)$);
\coordinate(c3) at ($(C3)+(0,-1)$);


\coordinate(v1) at (0.7,0-1);
\coordinate(v2) at (1.6,0.2-1);
\coordinate(v3) at (1.2,-0.25-1);
\coordinate(v4) at (2.35,-0.1-1);

\draw (v1) node {\blue{$\bullet$}};
\draw (v2) node {\blue{$\bullet$}};
\draw (v3) node {\blue{$\bullet$}};
\draw (v4) node {\blue{$\bullet$}};
\draw[thick,blue] (v1)--(v2)--(v4)--(v3)--(v1);
\draw[thick,blue] (v1)--(v4);

\coordinate(w1) at (0.7,4+0+1);
\coordinate(w2) at (2.4, 4+0.15+1);
\coordinate(w3) at (1.6,4+-0.25+1);
\draw (w1) node {\blue{$\bullet$}};
\draw (w2) node {\blue{$\bullet$}};
\draw (w3) node {\blue{$\bullet$}};
\draw[thick,blue] (w1)--(w2)--(w3)--(w1);

\draw[decoration={markings,mark=at position 0.6 with {\arrow[red,scale=1.5,>=stealth]{>}}},postaction={decorate}] (v1) --(b1);
\draw[decoration={markings,mark=at position 0.6 with {\arrow[red,scale=1.5,>=stealth]{>}}},postaction={decorate}] (v2)--(b2);
\draw[decoration={markings,mark=at position 0.6 with {\arrow[red,scale=1.5,>=stealth]{>}}},postaction={decorate}] (v3)--(b3);
\draw[decoration={markings,mark=at position 0.6 with {\arrow[red,scale=1.5,>=stealth]{>}}},postaction={decorate}] (v4)--(b4);

\draw[decoration={markings,mark=at position 0.6 with {\arrow[red,scale=1.5,>=stealth]{>}}},postaction={decorate}] (c1)--(w1);
\draw[decoration={markings,mark=at position 0.6 with {\arrow[red,scale=1.5,>=stealth]{>}}},postaction={decorate}] (c2)--(w2);
\draw[decoration={markings,mark=at position 0.6 with {\arrow[red,scale=1.5,>=stealth]{>}}},postaction={decorate}] (c3)--(w3);


\coordinate(x1) at (0.3,1.3);
\coordinate(x2) at (1.8,2.8);
\coordinate(x3) at (2.7,2.2);
\draw (x1) node {{$\bullet$}};
\draw (x2) node {{$\bullet$}};
\draw (x3) node {{$\bullet$}};
\draw(x1) to[in=-130,out=0,looseness=.7]  (x3);

\draw(c3)--(x3)--(b4);
\draw(x3)--(c2);
\draw[dotted] (b2)--(x2) ;
\draw(b3)--(x1)--(b1);
\draw[dotted] (c1) to[in=170,out=-90,looseness=.9]  (x2);
\draw (x1) to[in=-90,out=-170,looseness=.9] (0,1.5) ;
\draw[dotted] (0,1.5) to[in=-160,out=90,looseness=.9]  (x2);
\draw (x3) to[in=-90,out=60,looseness=.9] (3,3) ;
\draw[dotted] (3,3) to[in=30,out=90,looseness=.3]  (x2);

\node[left] at (Ai) {\blue{$\gamma_i$}};
\node[left] at (Af) {\blue{$\gamma_f$}};
\node[left] at (0,2) {{$\gamma$}};

\node[right] at (3,-.5) {\red{$\{H_{e}\}_{e\in\pp\gamma_{i}}$}};
\node[right] at (3,4.5) {\red{$\{H_{e}\}_{e\in\pp\gamma_{i}}$}};

\end{tikzpicture}

\caption{Gluing back the open spin networks on the initial and final disks to the cylinder spin networks with the corner data of $\SU(2)$ holonomies {\red$\{H_{e}\}_{e\in\pp\gamma_{i}}$} and {\red$\{H_{e}\}_{e\in\pp\gamma_{i}}$} along the open edges.}
\label{fig:openspinnetwork}
\end{figure}

We apply  this same definition of spin network basis states to the cylinder graph $\gamma$, as well as the final disks on the open graphs $\gamma_{f}$, by adapting the orientation of the boundary edges. Then, as illustrated on fig.\ref{fig:openspinnetwork}, the open spin networks on the initial and final disks are glued to the open spin network on the cylinder through the corner data of $\SU(2)$ holonomies along the boundary edges $e\in\pp\gamma$. These boundary $\SU(2)$ group elements $\{g_{e}\}_{e\in\pp\gamma}$ map boundary spin states $|j_{e},m_{e}\ra$ on the cylinder boundary $e\in\pp\gamma$ to boundary spin states $|j_{e},m_{e}\ra$ on the disks' boundary $e\in\pp\gamma_{i}\cup\pp\gamma_{f}$.

Let us now analyze the Ponzano-Regge amplitudes for 3d quantum gravity on the solid cylinder and how it depends on the boundary state.

\subsection{Ponzano-Regge transition amplitude}


As shown in \cite{Freidel:2005bb,Dowdall:2009eg,Dittrich:2018xuk}, the Ponzano-Regge amplitude for a 3-ball does not depend on the chosen bulk triangulation (or cellular complex) but only on the boundary graph and spin network state on the boundary 2-sphere. Considering the spin network state $\psi$ defined on the graph $\Gamma$ on the 2-sphere, the Ponzano-Regge amplitude is simply the evaluation of the wave-function on the flat connection\footnotemark, $g_{e}=\id$ for all edges $e\in\Gamma$:
\be
\cA^{PR}[\psi_{\Gamma}]
=
\psi(\{\id\}_{e\in\Gamma})
\,.
\ee
\footnotetext{
Since the first homotopy group of the 2-sphere is trivial, there exists a unique flat connection on the graph $\Gamma$ (i.e. such that the holonomy around every closed loop of the graph is equal to $\id$) up to gauge transformations. Since  spin network wave-functions are gauge-invariant by definition, their evaluation on flat connections is constant and always equal to their evaluation on the trivial connection $g_{e}=\id$ for all edges $e\in\Gamma$.
}
This amounts to the projection of the spin network state on the physical state, peaked on the unique flat connection of the 2-sphere.

\medskip


Let us now turn to the solid cylinder, seen as a 3-ball. Its 2-sphere boundary is to be split, as explained in the previous section, into the initial disk, final disk and 2d cylinder linking them, and this canonical splitting results in the decomposition of the spherical boundary graph $\Gamma$ into three open pieces, spanning the three surfaces:
\be
\cS_{2}=\cD_{i}\cup\cC\cup\cD_{f}
\,,\qquad
\Gamma= \gamma_{i}\# \gamma \#\gamma_{f}
\,,
\ee
where we use the $\#$ to mean that these open graphs are glued along their shared boundary edges, as illustrated on fig.\ref{fig:graphdecomposition}.
This  translates into the decomposition of the boundary spin network state similarly as:
\be
\psi=\psi_{i}\#\Psi\#\psi_{f},
\ee
where the gluing operation is done by a convolution product \eqref{glueg} in the representation of the states as wave-functions of $\SU(2)$ holonomies or by the matching of the boundary spin states up to corner holonomies \eqref{gluespin} in the spin basis.

Let us start with the formulation in terms of group elements. The Ponzano-Regge evaluation with boundary state $\psi_{i}\#\Psi\#\psi_{f}$ gives:
\be
\cA^{PR}[\psi_{i}\#\Psi\#\psi_{f}]
=
\int \prod_{e\in\pp\gamma}\rd H_{e}\,
\psi_{i}\big{(}\{\id\}_{\gamma^o_{i}}, \{H_{e}\}_{e\in\pp\gamma_{i}}\big{)}
\psi_{f}\big{(}\{\id\}_{\gamma^o_{f}}, \{H_{e}^{-1}\}_{e\in\pp\gamma_{f}}\big{)}
\,
\Psi\big{(}\{\id\}_{\gamma^o}, \{H_{e}^{-1}\}_{e\in\pp\gamma_{i}},\{H_{e}\}_{e\in\pp\gamma_{f}}\big{)}
\,,
\ee
where we have set all the $\SU(2)$ group elements living in the bulk of the disks and of the cylinder to the identity and integrate over the corner group elements in order to glue the three pieces together. This shows that the bulk of the disk states does not matter and the transition amplitude only depends on the state of the boundary of the disk. Let us introduce the notation for the boundary state of the initial disk:
\be
\psi^{\pp}_{i}\big{(} \{H_{e}\}_{e\in\pp\gamma_{i}}\big{)}
=
\psi_{i}\big{(}\{\id\}_{\gamma^o_{i}}, \{H_{e}\}_{e\in\pp\gamma_{i}}\big{)}
\,.
\ee
This boundary state only depends on the corner group elements, i.e. group elements living along the open links of the disk. It is gauge-invariant under the right $\SU(2)$ action, $\psi^{\pp}_{i}\big{(} \{H_{e}\}\big{)}=\psi^{\pp}_{i}\big{(} \{H_{e}h^{-1}\}\big{)}$ for all $h\in\SU(2)$. This means that the initial disk's boundary state is essentially an intertwiner state.
We introduce the same notation for the final disk:
\be
\psi^{\pp}_{f}\big{(} \{H_{e}\}_{e\in\pp\gamma_{f}}\big{)}
=
\psi_{f}\big{(}\{\id\}_{\gamma^o_{f}}, \{H_{e}\}_{e\in\pp\gamma_{f}}\big{)}
\,.
\ee
Then the cylindric Ponzano-Regge amplitude describing the evolution of the initial disk to the final disk through the solid cylinder only depends on the initial and final boundary states and the transition kernel is given by the cylinder spin network functional $\Psi$:
\be
\cA^{PR}[\psi_{i}\#\Psi\#\psi_{f}]
=
\int \prod_{e\in\pp\gamma_{i}}\rd H_{e}\, \prod_{e\in\pp\gamma_{f}}\rd \widetilde{H}_{e}\,
\psi^{\pp}_{i}\big{(} \{H_{e}\}\big{)}
\psi^{\pp}_{f}\big{(} \{\widetilde{H}_{e}^{{-1}}\}\big{)}
\,
\Psi\big{(}\{\id\}_{\gamma^o}, \{H_{e}^{-1}\},\{\widetilde{H}_{e}\}\big{)}
\,,
\ee
where we distinguished the group elements on the initial and final corners.

\medskip 

Switching to the spin basis allows for a more operational point of view. Let us consider the case of a spin network basis on the whole solid cylinder boundary, i.e. we fix the spins $j_{e}$ on all the edges of the boundary graph $\Gamma$, as well as the intertwiner states $I_{v}$ on its vertices. This assumption is to keep notations as simple as possible and all that follows can be easily adapted to superpositions of spins and intertwiners.

The initial disk boundary state, or initial corner state, is a spin state living in the tensor product of the spins attached to the open edges of the initial graph, i.e. the initial boundary Hilbert space is
\be
\cH^{\pp}_{i}=\bigotimes_{e\in\pp\gamma_{i}} \cV_{j_{e}}\,,
\ee
and similarly for the final boundary state:
\be
\cH^{\pp}_{f}=\bigotimes_{e\in\pp\gamma_{f}} \cV_{j_{e}}\,.
\ee
The cylindric Ponzano-Regge amplitude $\cA^{PR}_{\gamma}[\psi^{\pp}_{i},\psi^{\pp}_{f}]$ is then given by the evaluation of the cylinder spin network interpolating between the spins on the initial corner and the final corner state.. In the usual spin basis labeled by the magnetic indices, the Ponzano-Regge transition amplitude for the cylinder graph $\gamma$ reads:
\be
\la\bigotimes_{e\in\pp\gamma_{f}} j_{e},m_{e}^{t}|\cA^{PR}_{\gamma}|\bigotimes_{e\in\pp\gamma_{i}} j_{e},m_{e}^{s}\ra
=
\sum_{\{m_{e}^{s,t}\}_{e\in\gamma}}\prod_{v\in\gamma}
\la \bigotimes_{e|v=s(e)}j_{e},m^{s}_{e}\,|I_{v}|\,\bigotimes_{e|v=t(e)}j_{e},m^{t}_{e}\ra
\,,
\ee
where the initial and final corner states are here to close the open edges of the spin network state on the 2d cylinder.

This shows how the Ponzano-Regge model defines transition amplitudes for evolution of the 2d disk geometry, which turns out to depend solely on the boundary state of the 2d disk and the boundary spin network  encoding the quantum state of the 1+1-d  geometry between the initial and final disks.

The key property of those cylindric Ponzano-Regge amplitude with boundary is that they are invariant under the action of $\SU(2)$, in the sense that the boundary evaluation commutes with the  $\SU(2)$ action:
\be
\forall g\in\SU(2)\,\qquad
\bigg{(}\la\bigotimes_{e\in\pp\gamma_{f}} j_{e},m_{e}^{t}|g^{-1}\bigg{)}
\,|\cA^{PR}_{\gamma}|\,
\bigg{(}\bigotimes_{e\in\pp\gamma_{i}} g|j_{e},m_{e}^{s}\ra\bigg{)}
=
\la\bigotimes_{e\in\pp\gamma_{f}} j_{e},m_{e}^{t}|\cA^{PR}_{\gamma}|\bigotimes_{e\in\pp\gamma_{i}} j_{e},m_{e}^{s}\ra
\,,
\ee
as follows directly from its expression in terms of intertwiners given above. We will show in the next section that this implies that the total spin is preserved during the evolution along the cylinder and that this means that the Ponzano-Regge amplitude will dynamically select which total spin eigenmode(s) will dominate the evolution and survive at large time in a continuum limit. In the following, we will focus on a square lattice on the boundary, to have an obvious way to scale the boundary graph. But one should keep in time that the $\SU(2)$-invariance of the amplitudes is a universal property, which does not depend on the choice of boundary foliation.

The rest of this paper is dedicated to studying the structure and properties of this Ponzano-Regge transition amplitudes, in particular its reformulation in terms of transfer matrix in section \ref{sec:transfer}, and illustrates the construction through the simplest non-trivial examples, focusing on the evolution of one, two and three spin states in section \ref{sec:example}.

\section{Transfer matrix \& total spin eigenmodes}
\label{sec:transfer}

\subsection{Transfer matrix on the square lattice}

\begin{figure}[h!]

				\begin{tikzpicture}[scale=1.1]
						\coordinate(ai) at (0,0);
				\coordinate(bi) at (0,5);

				\draw[fill=orange, fill opacity=.3](0,0) to[in=-90,out=-90,looseness=.5]   node[pos=0.2,inner sep=0pt](a1){}  node[pos=0.4,inner sep=0pt](a2){}  node[pos=0.6,inner sep=0pt](a3){}  node[pos=0.8,inner sep=0pt](a4){} (3,0)to[in=90,out=90,looseness=.5] (0,0);

				\draw[fill=orange, fill opacity=.3](0,5) to[in=-90,out=-90,looseness=.5]   node[pos=0.2,inner sep=0pt](b1){}  node[pos=0.4,inner sep=0pt](b2){}  node[pos=0.6,inner sep=0pt](b3){}  node[pos=0.8,inner sep=0pt](b4){} (3,5)to[in=90,out=90,looseness=.5] (0,5);
								
				\coordinate(af) at (3,0);
				\coordinate(bf) at (3,5);
				
				\foreach \i in {1,...,4}{
					\draw[thick,in=-90,out=-90,looseness=.5] (0,\i) to node[pos=0.2,inner sep=0pt]{$\bullet$}  node[pos=0.4,inner sep=0pt]{$\bullet$}  node[pos=0.6,inner sep=0pt]{$\bullet$}  node[pos=0.8,inner sep=0pt]{$\bullet$} (3,\i);
					\draw[dotted,thick,in=90,out=90,looseness=.5] (0,\i) to (3,\i);
				}
						
\draw[thick] (ai) -- (bi);
\draw[thick,decoration={markings,mark=at position 0.13 with {\arrow[scale=1.5,>=stealth]{>}}},decoration={markings,mark=at position 0.93 with {\arrow[scale=1.5,>=stealth]{>}}},postaction={decorate}](a1) -- (b1);
				\draw[thick,decoration={markings,mark=at position 0.13 with {\arrow[scale=1.5,>=stealth]{>}}},decoration={markings,mark=at position 0.93 with {\arrow[scale=1.5,>=stealth]{>}}},postaction={decorate}](a2) --(b2);
				\draw[thick,decoration={markings,mark=at position 0.13 with {\arrow[scale=1.5,>=stealth]{>}}},decoration={markings,mark=at position 0.93 with {\arrow[scale=1.5,>=stealth]{>}}},postaction={decorate}](a3) --  (b3);
				\draw[thick,decoration={markings,mark=at position 0.13 with {\arrow[scale=1.5,>=stealth]{>}}},decoration={markings,mark=at position 0.93 with {\arrow[scale=1.5,>=stealth]{>}}},postaction={decorate}] (a4) --(b4);
				\draw[thick] (af) -- (bf);
\node at (1.5,0) {\blue{$\gamma_{i}$}};
\node at (1.5,5) {\blue{$\gamma_{f}$}};
\node[left] at (0,2.5) {{$\gamma$}};

\node at ($(a1)+(-.5,-.15)$) {{$(j_{1}^{0},m_{1}^{0})$}};
\node at ($(a2)+(-.15,-.2)$) {{$(j_{2}^{0},m_{2}^{0})$}};
\node at ($(a3)+(.15,-.2)$) {{$(j_{3}^{0},m_{3}^{0})$}};
\node at ($(a4)+(.5,-.15)$) {{$(j_{4}^{0},m_{4}^{0})$}};
				\end{tikzpicture}

\caption{The open spin network on the  2d cylinder living on a square lattice mapping the spin states on the initial disk boundary to the spin states on the final disk boundary.}
\label{fig:cylinderlattice}
\end{figure}
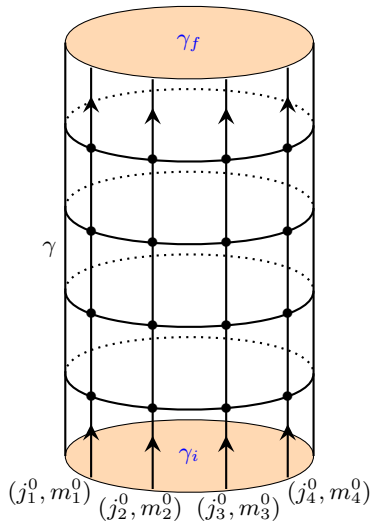

Let us study the simplest scalable cylindric configuration, where the graph on the 2d cylinder is a square lattice interpolating between initial and final states with the same number of boundary edges, as drawn on fig.\ref{fig:cylinderlattice}.
We call $N$ the number of initial (and thus of final) spins and $S$ the number of time slices. The square lattice on the boundary cylinder consists in $NS$ vertices, each carrying an intertwiner. 

\begin{figure}[h!]

\hspace*{-36mm}
\begin{tikzpicture}[scale=1.9]

\coordinate(v1) at (0,0) ;
\coordinate(v2) at (1,0) ;
\coordinate(v3) at (2,0);
\coordinate(v) at (3.5,0);
\coordinate(vN1) at (5,0);
\coordinate(vN) at (6,0);

\node at (v1) {$\bullet$};
\node at (v2) {$\bullet$};
\node at (v3) {$\bullet$};
\node at (v) {$\bullet$};
\node at (vN1) {$\bullet$};
\node at (vN) {$\bullet$};

\node[below left] at (v1) {$I_{1}^{\tau}$};
\node[below left]  at (v2) {$I_{2}^{\tau}$};
\node[below left]  at (v3) {$I_{3}^{\tau}$};
\node[below left]  at (v) {$I_{e}^{\tau}$};
\node[below left]  at (vN1) {$I_{N-1}^{\tau}$};
\node[below left]  at (vN) {$I_{N}^{\tau}$};

\draw[thick] (v1)++(-.5,0) -- ($(v3)+(.5,0)$);
\draw[thick] (vN1)++(-.5,0) -- ($(vN)+(.5,0)$);
\draw[thick] ($(v)+(-.5,0)$) -- ($(v)+(.5,0)$);
\draw[thick,dotted] ($(v3)+(.5,0)$) -- ($(v)+(-.5,0)$);
\draw[thick,dotted] ($(v)+(.5,0)$) -- ($(vN1)+(-.5,0)$);

\draw[thick,decoration={markings,mark=at position 0.3 with {\arrow[scale=1,>=stealth]{>}}},decoration={markings,mark=at position 0.85 with {\arrow[scale=1,>=stealth]{>}}},postaction={decorate}] ($(v)+(0,-.5)$)node[below]{$(j_{e}^{\tau-1},m_{e}^{\tau-1})$}--($(v)+(0,.5)$)node[above]{$(j_{e}^{\tau},m_{e}^{\tau})$};
\draw[thick,decoration={markings,mark=at position 0.3 with {\arrow[scale=1,>=stealth]{>}}},decoration={markings,mark=at position 0.85 with {\arrow[scale=1,>=stealth]{>}}},postaction={decorate}] ($(v1)+(0,-.5)$) node[below]{$(j_{1}^{\tau-1},m_{1}^{\tau-1})$}--($(v1)+(0,.5)$)node[above]{$(j_{1}^{\tau},m_{1}^{\tau})$};
\draw[thick,decoration={markings,mark=at position 0.3 with {\arrow[scale=1,>=stealth]{>}}},decoration={markings,mark=at position 0.85 with {\arrow[scale=1,>=stealth]{>}}},postaction={decorate}] ($(v2)+(0,-.5)$)--($(v2)+(0,.5)$);
\draw[thick,decoration={markings,mark=at position 0.3 with {\arrow[scale=1,>=stealth]{>}}},decoration={markings,mark=at position 0.85 with {\arrow[scale=1,>=stealth]{>}}},postaction={decorate}] ($(v3)+(0,-.5)$)--($(v3)+(0,.5)$);
\draw[thick,decoration={markings,mark=at position 0.3 with {\arrow[scale=1,>=stealth]{>}}},decoration={markings,mark=at position 0.85 with {\arrow[scale=1,>=stealth]{>}}},postaction={decorate}] ($(vN1)+(0,-.5)$)--($(vN1)+(0,.5)$);
\draw[thick,decoration={markings,mark=at position 0.3 with {\arrow[scale=1,>=stealth]{>}}},decoration={markings,mark=at position 0.85 with {\arrow[scale=1,>=stealth]{>}}},postaction={decorate}] ($(vN)+(0,-.5)$)node[below]{$(j_{N}^{\tau-1},m_{N}^{\tau-1})$}--($(vN)+(0,.5)$)node[above]{$(j_{N}^{\tau},m_{N}^{\tau})$};

\draw[dotted,in=20,out=160,looseness=1.2] ($(v1)+(-.5,0)$) to  ($(vN)+(.5,0)$);

\draw[thick,decoration={markings,mark=at position 0.5 with {\arrow[scale=1,>=stealth]{>}}},postaction={decorate}] (v1)-- (v2);
\node at (.5,0.15){$k_{1}^{\tau},n_{1}^{\tau}$};
\draw[thick,decoration={markings,mark=at position 0.5 with {\arrow[scale=1,>=stealth]{>}}},postaction={decorate}] (v2)-- (v3);
\node at (1.5,0.15){$k_{2}^{\tau},n_{2}^{\tau}$};
\draw[thick,decoration={markings,mark=at position 0.5 with {\arrow[scale=1,>=stealth]{>}}},postaction={decorate}] (vN1)-- (vN);
\node at (5.5,0.15){$k_{N-1}^{\tau},n_{N-1}^{\tau}$};
\draw[thick,decoration={markings,mark=at position 0.2 with {\arrow[scale=1,>=stealth]{>}}},postaction={decorate}] ($(v)+(-.5,0)$)-- (v);
\node at (3,0.15){$k_{e-1}^{\tau},n_{e-1}^{\tau}$};
\draw[thick,decoration={markings,mark=at position .9 with {\arrow[scale=1,>=stealth]{>}}},postaction={decorate}] (v)--($(v)+(.5,0)$);
\node at (4,0.15){$k_{e}^{\tau},n_{e}^{\tau}$};
\draw[thick,decoration={markings,mark=at position 0.5 with {\arrow[scale=1,>=stealth]{>}}},postaction={decorate}] (v1)-- (v2);
\node at (.5,0.15){$k_{1}^{\tau},n_{1}^{\tau}$};
\draw[thick,decoration={markings,mark=at position 0.9 with {\arrow[scale=1,>=stealth]{>}}},postaction={decorate}] (vN)--($(vN)+(.5,0)$);
\node at (6.5,0.15){$k_{N}^{\tau},n_{N}^{\tau}$};

\end{tikzpicture}

\caption{A single time slice of the spin network on the boundary cylinder as a concatenation of $N$ intertwiners mapping the incoming  spin states $\otimes_{e}|j_{e}^{\tau-1},m_{e}^{\tau-1}\ra$ onto outgoing spin states $\otimes_{e}|j_{e}^{\tau},m_{e}^{\tau}\ra$.}
\label{fig:singleslice}
\end{figure}
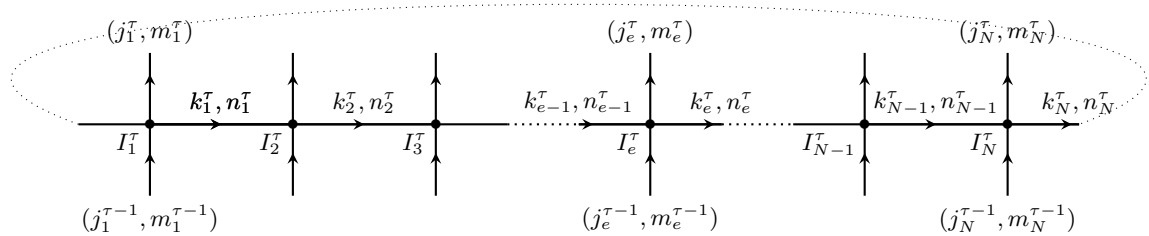

Let us start with the case of a single time slice $S=1$. The $N$ incoming spin states $|j_{e}^{0},m_{e}^{0}\ra$, with the label $e$ running from 1 to $N$, each connect to a 4-valent intertwiner and those intertwiners form a closed loop (which goes around the cylinder). In simpler terms, the $N$ incoming spins all recouple with a spin travelling on the transverse loop.
We write $v_{e}^{1}$ for the vertex connected to the open edge $e$, with the superscript $1$ referring to the time slice. We similarly call $I_{e}^{1}$ the intertwiner living at the vertex  $v_{e}^{1}$. We call $k_{e}^{1}$ the spin colouring the edge linking the vertex $v_{e}^{1}$ to the vertex $v_{e+1}^{1}$, with the implicit convention that the index $e$ is cyclic, i.e. $e=N+1\equiv1\,\textrm{mod}\,N$. Then the intertwiner $I_{e}^{1}$ recouples the incoming spin $j_{e}^{0}$ with the tranverse spin $k_{e-1}^{1}$ into the outgoing spin $j_{e}^{1}$ and the transverse spin $k_{e}^{1}$, following the orientation of fig.\ref{fig:singleslice},
\be
I_{e}^{1}:\cV_{j_{e}^{0}}\otimes\cV_{k_{e-1}^{1}}\longrightarrow \cV_{j_{e}^{1}}\otimes\cV_{k_{e}^{1}}\,.
\ee

The transition amplitude between the incoming spin states and outgoing spin states is then given by the tensor product of the intertwiners $\bigotimes_{e}I_{e}^{1}$ traced over the transverse spins:
\beq
\la\{ j_{e}^{1},m_{e}^{1}\}|T_{1}|\{j_{e}^{0},m_{e}^{0}\}\ra
&=&
\la\{ j_{e}^{1},m_{e}^{1}\}|
\tr_{\{k_{e}^{1}\}}\bigotimes_{e}I_{e}^{1}
|\{j_{e}^{0},m_{e}^{0}\}\ra
\nn\\
&=&
\sum_{n_{e}}
\prod_{e=1}^{N}
\la (j_{e}^{1},m_{e}^{1})(k_{e}^{1},n_{e})\,|\,I_{e}^{1}\,|\,(j_{e}^{0},m_{e}^{0})(k_{e-1}^{1},n_{e-1})\ra\,.
\eeq
This defines the map $T_{1}$ between the initial corner spin states and the final corner spin states defined by a single slice of intertwiners on the boundary cylinder.
It allows to recast the Ponzano-Regge transition amplitude in terms of a transfer matrix defined through a partial trace of the tensor product of local evolution maps, as in condensed matter and (integrable) spin chains.
Iterating this formalism, we define a transfer matrix $T_{\tau}$ for each time slice $\tau=1..S$ with the overall evolution between the initial and final corner given by the composed map:
\be
\cT=T_{S}\circ..\circ T_{1}\,
:\bigotimes_{e\in\pp\gamma_{i}} \cV_{j_{e}}\longrightarrow\bigotimes_{e\in\pp\gamma_{f}} \cV_{j_{e}}
\,.
\ee
This is the boundary evolution map defined by the  Ponzano-Regge path integral for 3d quantum geometry applied to a solid cylinder with a boundary spin network.

\subsection{Total spin selection in the refinement limit}

The main property of the transfer matrix is that it is $\SU(2)$-invariant, or more precisely that it commutes with the $\SU(2)$ action:
\be
\forall \tau\,,\,\,\forall h\in\SU(2)\qquad 
T_{\tau}\circ D(h)=D(h)\circ T_{\tau}\,,
\ee
where $D(h)$ denotes the action of the $\SU(2)$ group element $h$ either on the initial spin states (on the left hand side) or on the final spin states (on the right hand side). This is a straightforward consequence of the local gauge invariance under $\SU(2)$ transformations of the Ponzano-Regge path integral, which results into the Ponzano-Regge amplitude given by the boundary spin network evaluation defined as the direct gluing of intertwiners on the boundary without any intermediate non-trivial $\SU(2)$ holonomy in-between.

Since this is true for each time slice, it is also automatically valid for the overall evolution map, $\cT\circ D(h)=D(h)\circ \cT$. In practice, this means that the overall recoupled spin is conserved throughout the evolution.
Let us make this explicit and introduce the recoupled basis for the $N$ incoming spins. Considering a single spin $j\in\f\N2$ defining an irreducible representation of the $\SU(2)$ group acting on the Hilbert space $\cV_{j}$, we call the $\su(2)$ generators $J_{a}$ with $a=1..3$ and write them as a vector operator $\vJ$, with the standard commutation relations:
\be
[J_{a},J_{b}]=i\eps_{abc}J_{c}\,.
\ee
The magnetic basis $|j,m\ra$, with $m$ running from $-j$ to $+j$ by integer step, diagonalizes one of the $\su(2)$ generators, say $J_{3}$ following the usual conventions, and the quadratic Casimir:
\be
\vJ^{2}|j,m\ra=j(j+1)|j,m\ra\,,\qquad
J_{3}|j,m\ra=m|j,m\ra\,.
\ee
Now considering the tensor product of several spins $j_{e}$ with $e=1..N$, we look at the simultaneous diagonal action of $\SU(2)$ on the overall representation:
\be
\forall h\in\SU(2)\,,\quad
h\triangleright\bigotimes_{e}|j_{e},m_{e}\ra
=
\bigotimes_{e}D^{j_{e}}(h)\,|j_{e},m_{e}\ra\,,\qquad
\vJ\,
\bigotimes_{e}|j_{e},m_{e}\ra
=
\sum_{\tilde{e}}\vJ^{(\tilde{e})}
\bigotimes_{e}|j_{e},m_{e}\ra\,,
\ee 
where each Wigner matrix $D^{j_{e}}(h)$ acts on its corresponding spin $j_{e}$ and the $\su(2)$ generators $\vJ^{({e})}$ acts on the Hilbert space $\cV_{j_{e}}$.
The recoupled basis diagonalizes the Casimir operator of this overall $\SU(2)$ action:
\be
\bigotimes_{e}\cV_{j_{e}}
=
\bigoplus_{J\in\f\N2}\cV_{J}\otimes \cN_{J}
\,,\qquad\textrm{with the multiplicity  spaces}\quad
\cN_{J}\equiv\textrm{Inv}_{\SU(2)}\big{[}\cV_{J}\otimes\bigotimes_{e}\cV_{j_{e}}\big{]}\,.
\ee
The multiplicity spaces are intertwiner spaces between the individual spins $j_{e}$ and the overall recoupled spin $J$. Their dimension are given as integrals over $\SU(2)$ of products of characters:
\be
\dim\cN_{J}=\int_{\SU(2)}\rd g\,\chi_{J}(g)\prod_{e}\chi_{j_{e}}(g)\,,
\ee
where the character in a representation ofd spin $j$ is the trace of the Wigner matrices, $\chi_{j}(g)=\tr D^{j}(g)$.
A consistency check is given by computing the dimension of the tensor product of the $N$ spins:
\beq
\sum_{J}\dim\cV_{J}\dim\cN_{J}
&=&
\sum_{J}(2J+1)\dim\cN_{J}\nn\\
&=&
\int_{\SU(2)}\rd g\,\Big{[}\sum_{J}(2J+1)\chi_{J}(g)\Big{]}\prod_{e}\chi_{j_{e}}(g)
=
\int_{\SU(2)}\rd g\,\delta(g)\prod_{e}\chi_{j_{e}}(g)\nn\\
&=&
\prod_{e}(2j_{e}+1)
=
\dim\bigotimes_{e}\cV_{j_{e}}
\,.
\nn
\eeq
We introduce a basis $\cI^{(J)}_{\alpha}$ for the multiplicity space $\cN_{J}$ and perform a change of basis from the decoupled spins to the recoupled spin:
\be
\left|\begin{array}{rlcrl}
(\vJ^{(e)})^{2}&\bigotimes_{e}|j_{e},m_{e}\ra&=&j_{e}(j_{e}+1)&\bigotimes_{e}|j_{e},m_{e}\ra\,,
\vspace*{1mm}\\
J^{(e)}_{3}&\bigotimes_{e}|j_{e},m_{e}\ra
&=&
m_{e}&\bigotimes_{e}|j_{e},m_{e}\ra\,,
\end{array}
\right.
\quad{\Large\rightsquigarrow}\quad
\left|\begin{array}{rlcrl}
(\vJ^{(e)})^{2}&|\{j_{e}\},J,M,\cI^{(J)}_{\alpha}\ra
&=&
j_{e}(j_{e}+1)&|\{j_{e}\},J,M,\cI^{(J)}_{\alpha}\ra\,,
\vspace*{1mm}\\
(\vJ)^{2}&|\{j_{e}\},J,M,\cI^{(J)}_{\alpha}\ra&=&J(J+1)&|\{j_{e}\},J,M,\cI^{(J)}_{\alpha}\ra\,,
\vspace*{1mm}\\
J_{3}&|\{j_{e}\},J,M,\cI^{(J)}_{\alpha}\ra
&=&
M&|\{j_{e}\},J,M,\cI^{(J)}_{\alpha}\ra\,.
\end{array}
\right.
\nn
\ee

\medskip

The $\SU(2)$ covariance of the transfer matrix of each time slice means that it commutes with the recoupled spin operators, i.e. that the recoupled spin $J$ and magnetic moment $M$ are conserved throughout the evolution:
\be
\Big{[}
T_{\tau},(\vJ)^{2}
\Big{]}
=
\Big{[}
T_{\tau},J_{3}
\Big{]}
=
0\,,
\qquad
\left|\begin{array}{rlcrl}
(\vJ)^{2}\,T_{\tau}\,
&|\{j_{e}\},J,M,\cI^{(J)}_{\alpha}\ra
&=&
J(J+1)\,T_{\tau}\,
&|\{j_{e}\},J,M,\cI^{(J)}_{\alpha}\ra
\,,
\vspace*{1mm}\\
J_{3}\,T_{\tau}\,
&|\{j_{e}\},J,M,\cI^{(J)}_{\alpha}\ra
&=&
M\,T_{\tau}\,
&|\{j_{e}\},J,M,\cI^{(J)}_{\alpha}\ra
\,.
\end{array}
\right.
\ee

Let us now consider the special case where the individual spins $j_{e}$ on the disk boundary are kept constant. This is a specific choice of (quantum) boundary condition where the corner is always made of the same number $N$ of 1d line elements with fixed quantized lengths $j_{e}$ (in Planck units), but whose geometry can still vary and evolve depending on the relative angles of those boundary line elements. Then the corner Hilbert space does not change:
\be
\cH^{\pp}_{i}=\cH^{\pp}_{f}=\bigotimes_{e=1}^{N} \cV_{j_{e}}=\cH^{\pp}\,.
\ee
Such boundary condition allow us to consider a homogeneous boundary ansatz, where every time slice are given by the same intertwiners. This means that the transfer of every time slice is the same and that the overall evolution map is given by a power of this one-slice transfer matrix:
\be
\forall \tau\,,\quad T_{\tau}=T\,,\qquad
\cT=T^{S}\,.
\ee
Since $T$ commutes with the recoupled spin operators $\vJ$, we decompose the corner Hilbert space as above, $\cH^{\pp}=\bigoplus_{J}\cV_{J}\otimes\cN_{J}$. Then the transfer matrix $T$ acts ``by block'' independently on each multiplicity space $\cN_{J}$.  Its eigenvalues thus depends on the basis labels $J,M$ and possibly some label $\beta$ within each space $\cN_{J}$. Writing them $\lambda_{J,M}^{(\beta)}$, the overall evolution map $\cT$ after $S$ time slices will have as eigenvalues $(\lambda_{J,M}^{(\beta)})^{S}$. 

This naturally results in a dynamical selection of the recoupled spin $(J,M)$ given by the highest eigenvalue (in modulus). The eigenvalues are determined by the choice of intertwiners defining the iterated time slice. Depending on that choice, we could select the highest possible recoupled spin $J_{max}=\sum_{e}j_{e}$, or we could project asymptotically, at large times $S\rightarrow\infty$, on the vanishing recoupled spin $J_{min}=0$. This latter case amounts to the dynamical selection of the $\SU(2)$-invariant subspace of boundary states, which could be interpreted as an emergent gauge symmetry. Depending on the choice of transfer matrix, we could also have a variety of intermediate cases between those two extreme possibilities with dynamical selection of some other recoupled spin in-between the minimal and maximal values, as well as possibly more spread limit probability distributions.
We illustrate this process in the next section with explicit examples for $N=1$, $N=2$ and $N=3$ spin states on the corner.

We conclude this section with the remark that, although we have fixed the spins and assume that they do not change, $j_{e}^{i}=j_{e}^{f}=j_{e}$, this assumption was only for the sake of keeping notations as simple as possible. There is absolutely no obstacle in considering spin superpositions and spin transitions. One would consider the (much) larger corner Hilbert space,
\be
\cH^{\pp}=\bigotimes_{e=1}^{N} \cH_{e}\,,\qquad
\cH_{e}=\bigoplus_{j_{e}}\cV_{j_{e}}\,,
\ee
decompose it into the recoupled basis $\cH^{\pp}=\bigoplus_{J}\cV_{J}\otimes\cN_{J}$ and study the action by block of the transfer matrices $T_{\tau}$ on each multiplicity space $\cN_{J}$ as above.

\section{Evolution on the cylinder}
\label{sec:example}

\subsection{(Trivial) evolution of the one-spin disk}

\begin{figure}[!]

\begin{tikzpicture}[scale=.8]
\coordinate(ai) at (0,0);
\coordinate(af) at (0,5);
\coordinate(bi) at (3,0);
\coordinate(bf) at (3,5);
				
\draw[fill=orange, fill opacity=.3,opacity=.3](ai) to[in=-90,out=-90,looseness=.5]   node[pos=0.5,inner sep=0pt](ci){} (bi)to[in=90,out=90,looseness=.5] (ai);

\draw[fill=orange, fill opacity=.3,opacity=.3](af) to[in=-90,out=-90,looseness=.5]   node[pos=0.5,inner sep=0pt](cf){} (bf)to[in=90,out=90,looseness=.5] (af);
				
\foreach \i in {1,...,4}{
\draw[thick](0,\i) to[in=-90,out=-90,looseness=.5]  node[pos=0.5,inner sep=0pt]{$\bullet$}  (3,\i);
\draw[dotted,thick,in=90,out=90,looseness=.5] (0,\i) to (3,\i);
}
						

\node at (ci) {$\red{\bullet}$};
\node at (cf) {$\red{\bullet}$};
\node at ($(ci)+(0,-.3)$) {{$(j^{0},m^{0})$}};
\draw[thick,decoration={markings,mark=at position 0.1 with {\arrow[scale=1.5,>=stealth]{>}}},decoration={markings,mark=at position 0.3 with {\arrow[scale=1.5,>=stealth]{>}}},decoration={markings,mark=at position 0.5 with {\arrow[scale=1.5,>=stealth]{>}}},decoration={markings,mark=at position 0.7 with {\arrow[scale=1.5,>=stealth]{>}}},decoration={markings,mark=at position 0.9 with {\arrow[scale=1.5,>=stealth]{>}}},postaction={decorate}] (ci)--(cf);
\end{tikzpicture}

\caption{Evolution of a single spin state traveling through a sequence of 4-valent intertwiners on the boundary cylinder.}
\label{fig:N1evolution}
\end{figure}
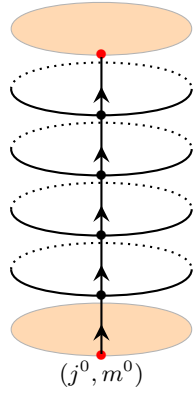

Let us consider the simplest case with a single spin on the corner, $N=1$. Geometrically, this corresponds to the disk boundary consisting in a single line element with quantized length. Let us call the corner spin $j$, dropping the $e$ index. The square lattice reduces to a central bone along the 2d boundary cylinder with loops defining the time slices, as drawn on fig.\ref{fig:N1evolution}. The spin $j$ will go through a sequence of $S$ 4-valent intertwiners.
As we show below, closing the transverse spins into loops implies that those 4-valent nodes do not affect the evolving spin at all. This is the direct expression of the $\SU(2)$ invariance of the transfer matrix in the single insertion $N=1$ on the disk boundary.

\begin{figure}[h!]

\begin{tikzpicture}[scale=1]

\draw[thick](0,0) to[in=-90,out=-90,looseness=.5]  node[pos=0.5,inner sep=0pt](O){$\bullet$}  (3,0);
\draw[dotted,thick,in=90,out=90,looseness=.5] (0,0) to (3,0);

\draw[thick,decoration={markings,mark=at position 0.3 with {\arrow[scale=1.5,>=stealth]{>}}},decoration={markings,mark=at position 0.8 with {\arrow[scale=1.5,>=stealth]{>}}},postaction={decorate}] ($(O)+(0,-1)$)node[below]{$|j,m\ra$}--($(O)+(0,1)$)node[above]{$|\tj,\tm\ra$};
\node[below left] at (O) {$I_{v}$};

\end{tikzpicture}

\caption{Closing the 4-intertwiner with a loop.}
\label{fig:intertwinerconvention}
\end{figure}
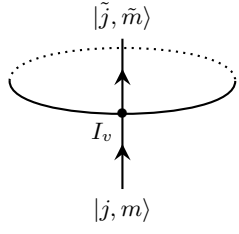

Indeed, let us start with an incoming spin state in $\cV_{j}$ evolving through an intertwiner $I_{v}$ into an outgoing spin spin state in $\cV_{\tj}$. As drawn on fig.\ref{fig:intertwinerconvention}, the transverse links of the intertwiner close to form a loop. Let us call $k$ the spin carried by that loop. The intertwiner $I_{v}$ is then a $\SU(2)$-invariant map from $\cV_{j}\otimes\cV_{k}$ to $\cV_{\tj}\otimes\cV_{k}$. The transfer matrix $T$ from $\cV_{j}$ to $\cV_{\tj}$ is derived by closing the loop, which amounts to taking the partial trace over the spin $k$:
\be
\la \tj,\tm|T|j,m\ra=\sum_{-k\le n\le +k}
\la (\tj,\tm)(k,n)|I_{v}|(j,m)(k,n)\ra
\,.
\ee

This transfer matrix then commutes with the  $\SU(2)$ action on the spin states:
\beq
\la \tj,\tm|\,h^{-1}Th\,|j,m\ra
&=&
\sum_n
\big{(}\la \tj,\tm|h^{-1}\otimes \la k,n|\big{)}
\,
|\,I_{v}\,|
\,
\big{(}h|j,m\ra\otimes |k,n\ra\big{)}\,,
\nn\\
&=&
\sum_n
\big{(}\la \tj,\tm|\otimes \la k,n|h\big{)}
\,
|\,I_{v}\,|
\,
\big{(}|j,m\ra\otimes h^{-1}|k,n\ra\big{)}\,,
\nn\\
&=&
\sum_n
\big{(}\la \tj,\tm|\otimes \la k,n|\big{)}
\,
|\,I_{v}\,|
\,
\big{(}|j,m\ra\otimes |k,n\ra\big{)}\,,
=\la \tj,\tm|\,T\,|j,m\ra
\,,
\eeq
where we used, first, the $\SU(2)$-invariance of the intertwiner and, second, the $\SU(2)$-invariance of the trace over the Hilbert space $\cV_{k}$.
$T$ can thus be understood as a bivalent intertwiner, i.e. a $\SU(2)$-invariant map from $\cV_{j}$ to $\cV_{\tj}$. This imposes that the incoming and outgoing spin states match\footnotemark{}, i.e. $j=\tj$. and $m=\tm$.
\footnotetext{
This mathematical fact can also be re-derived using the orthonormality of the Wigner matrices:
\be
\la \tj,\tm|\,T\,|j,m\ra
=
\la \tj,\tm|\,h^{-1}Th\,|j,m\ra
=
\int \rd h\,\la \tj,\tm|\,h^{-1}Th\,|j,m\ra
=
\la \tj,\tn|\,T\,|j,n\ra\,\int \rd h\,D^{\tj}_{\tm \tn}(h^{-1})D^{j}_{nm}(h)
=
\f{\delta_{\tj,j}}{2j+1}\,\delta_{\tn,n}\delta_{\tm,m}\la \tj,\tn|\,T\,|j,n\ra
\,.\nn
\ee
}
Thus, up to a normalization of the intertwiner and correspondingly of the transfer matrix, the evolution for a single disk insertion is trivial.

\subsection{The two spin$\f12$ propagator}

\begin{figure}[h!]
\begin{subfigure}{0.3\linewidth}
\begin{tikzpicture}[scale=.8]
\coordinate(ai) at (0,0);
\coordinate(af) at (0,5);
\coordinate(bi) at (3,0);
\coordinate(bf) at (3,5);
				
\draw[fill=orange, fill opacity=.3,opacity=.3](ai) to[in=-90,out=-90,looseness=.5]   node[pos=0.32,inner sep=0pt](ci){}node[pos=0.67,inner sep=0pt](di){} (bi)to[in=90,out=90,looseness=.5] (ai);

\draw[fill=orange, fill opacity=.3,opacity=.3](af) to[in=-90,out=-90,looseness=.5]   node[pos=0.32,inner sep=0pt](cf){}node[pos=0.67,inner sep=0pt](df){}(bf)to[in=90,out=90,looseness=.5] (af);
				
\foreach \i in {1,...,4}{
\draw[thick](0,\i) to[in=-90,out=-90,looseness=.5]  node[pos=0.32,inner sep=0pt]{$\bullet$}   node[pos=0.67,inner sep=0pt]{$\bullet$}  (3,\i);
\draw[dotted,thick,in=90,out=90,looseness=.5] (0,\i) to (3,\i);
}
						

\node at (ci) {$\red{\bullet}$};
\node at (cf) {$\red{\bullet}$};
\node at (di) {$\red{\bullet}$};
\node at (df) {$\red{\bullet}$};
\draw[thick,decoration={markings,mark=at position 0.1 with {\arrow[scale=1.5,>=stealth]{>}}},decoration={markings,mark=at position 0.3 with {\arrow[scale=1.5,>=stealth]{>}}},decoration={markings,mark=at position 0.5 with {\arrow[scale=1.5,>=stealth]{>}}},decoration={markings,mark=at position 0.7 with {\arrow[scale=1.5,>=stealth]{>}}},decoration={markings,mark=at position 0.9 with {\arrow[scale=1.5,>=stealth]{>}}},postaction={decorate}] (ci)--(cf);
\draw[thick,decoration={markings,mark=at position 0.1 with {\arrow[scale=1.5,>=stealth]{>}}},decoration={markings,mark=at position 0.3 with {\arrow[scale=1.5,>=stealth]{>}}},decoration={markings,mark=at position 0.5 with {\arrow[scale=1.5,>=stealth]{>}}},decoration={markings,mark=at position 0.7 with {\arrow[scale=1.5,>=stealth]{>}}},decoration={markings,mark=at position 0.9 with {\arrow[scale=1.5,>=stealth]{>}}},postaction={decorate}] (di)--(df);
\end{tikzpicture}
\end{subfigure}
\hspace*{10mm}
\begin{subfigure}{0.6\linewidth}
\begin{tikzpicture}[scale=1]

\draw[thick,decoration={markings,mark=at position 0.48 with {\arrow[scale=1,>=stealth]{>}}},decoration={markings,mark=at position .93 with {\arrow[scale=1,>=stealth]{>}}},postaction={decorate}](0,0) to[in=-90,out=-90,looseness=.5]  node[pos=0.32,inner sep=0pt](a){$\bullet$} node[pos=0.5,inner sep=1mm,above]{$k_{1}=\f12$} node[pos=0.67,inner sep=0pt](b){$\bullet$}node[pos=0.9,inner sep=0pt,below right]{$k_{2}=\f12$}  (3,0);
\draw[dotted,thick,in=90,out=90,looseness=.5] (0,0) to (3,0);

\draw[thick,decoration={markings,mark=at position 0.3 with {\arrow[scale=1,>=stealth]{>}}},decoration={markings,mark=at position 0.8 with {\arrow[scale=1,>=stealth]{>}}},postaction={decorate}] ($(a)+(0,-1)$)node[below]{$j_{1}={\f12}$}--($(a)+(0,1)$)node[above]{$\tj_{1}=\f12$};
\node[below left] at (a) {$I_{\alpha}$};
\draw[thick,decoration={markings,mark=at position 0.3 with {\arrow[scale=1,>=stealth]{>}}},decoration={markings,mark=at position 0.8 with {\arrow[scale=1,>=stealth]{>}}},postaction={decorate}] ($(b)+(0,-1)$)node[below]{$j_{2}={\f12}$}--($(b)+(0,1)$)node[above]{$\tj_{2}={\f12}$};
\node[below right] at (b) {$I_{\beta}$};

\node at (4.5,-.5) {\Large{=}};

\coordinate(x1) at  ($(a)+(5,-1)$);
\coordinate(y1) at  ($(a)+(5,1)$);
\coordinate(z1) at  ($(a)+(5,-0.4)$);
\coordinate(t1) at  ($(a)+(5,.4)$);
\coordinate(x2) at  ($(b)+(5,-1)$);
\coordinate(y2) at  ($(b)+(5,1)$);
\coordinate(z2) at  ($(b)+(5,-0.4)$);
\coordinate(t2) at  ($(b)+(5,.4)$);
\coordinate(z0) at  ($(z1)+(-.5,0)$);
\coordinate(t0) at  ($(t1)+(-.5,0)$);
\coordinate(z3) at  ($(z2)+(.5,0)$);
\coordinate(t3) at  ($(t2)+(.5,0)$);

\draw[thick,decoration={markings,mark=at position 0.5 with {\arrow[scale=1,>=stealth]{>}}},postaction={decorate}] (x1)--(z1);
\draw[thick,decoration={markings,mark=at position 0.6 with {\arrow[scale=1,>=stealth]{>}}},postaction={decorate}] (t1)--(y1);
\draw[thick,decoration={markings,mark=at position 0.5 with {\arrow[scale=1,>=stealth]{>}}},postaction={decorate}] (x2)--(z2);
\draw[thick,decoration={markings,mark=at position 0.6 with {\arrow[scale=1,>=stealth]{>}}},postaction={decorate}] (t2)--(y2);
\draw[fill=blue, fill opacity=.1](z0)--(z3)--(t3)--(t0)--(z0);
\node at ($.5*(a)+.5*(b)+(5,0)$) {2-qubit gate};
\node[below] at (x1) {$\cV_{\f12}$};
\node[below] at (x2) {$\cV_{\f12}$};
\node[above] at (y1) {$\cV_{\f12}$};
\node[above] at (y2) {$\cV_{\f12}$};

\end{tikzpicture}
\end{subfigure}
\caption{A pair of spin states evolving across a square lattice on the 2d cylinder.}
\label{fig:N2evolution}
\end{figure}
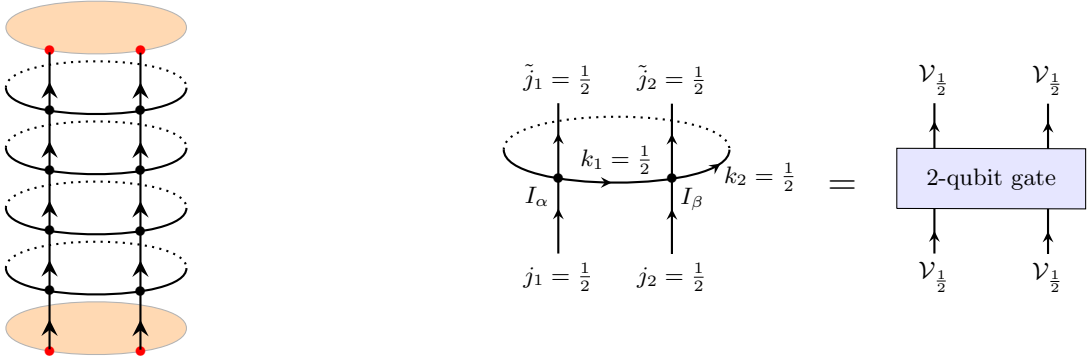

Let us illustrate the Ponzano-Regge transition amplitude with time-like boundary in the simplest, yet non-trivial, setting with a pair of incoming spin states. This case with $N=2$ disk insertions defines an evolution with time slices consisting of two incoming spin states getting recoupled through a pair of 4-valent intertwiners into two outgoing spin states, as drawn on fig.\ref{fig:N2evolution}.

\begin{figure}[h!]

\begin{subfigure}[t]{0.15\linewidth}

\begin{tikzpicture}[scale=0.6]

\draw[thick,decoration={markings,mark=at position 0.3 with {\arrow[scale=1,>=stealth]{>}}},decoration={markings,mark=at position 0.85 with {\arrow[scale=1,>=stealth]{>}}},postaction={decorate}] (-1,0) node[left]{${2}$}--(1,0)node[right]{${4}$};
\draw[thick,decoration={markings,mark=at position 0.3 with {\arrow[scale=1,>=stealth]{>}}},decoration={markings,mark=at position 0.85 with {\arrow[scale=1,>=stealth]{>}}},postaction={decorate}] (0,-1)node[below]{${1}$}--(0,1)node[above]{${3}$};

\node at (0,0) {$\bullet$};

\end{tikzpicture}

\caption{\label{fig:2qubita}
Incoming and outgoing spins}

\end{subfigure}
\hspace*{7mm}
\begin{subfigure}[t]{0.2\linewidth}

\begin{tikzpicture}[scale=0.6]

\draw[thick,decoration={markings,mark=at position 0.6 with {\arrow[scale=1,>=stealth]{>}}},postaction={decorate}] (-1,0) node[left]{${2}$}--(0,0);
\draw[thick,decoration={markings,mark=at position 0.6 with {\arrow[scale=1,>=stealth]{>}}},postaction={decorate}] (0,-1)node[below]{${1}$}--(0,0);
\draw[thick,decoration={markings,mark=at position 0.6 with {\arrow[scale=1,>=stealth]{>}}},postaction={decorate}] (1,1)--(2,1)node[right]{${4}$};
\draw[thick,decoration={markings,mark=at position 0.6 with {\arrow[scale=1,>=stealth]{>}}},postaction={decorate}] (1,1)--(1,2)node[above]{${3}$};

\node at (0,0) {$\bullet$};
\node at (1,1) {$\bullet$};

\draw[thick,red] (0,0)--(1,1);
\node at (1.5,.3) {\red\small $l=0,1$};

\end{tikzpicture}

\caption{\label{fig:4spinintertwiner}
Orthogonal spin basis of 4-valent inertwiners}

\end{subfigure}
\hspace*{2mm}
\begin{subfigure}[t]{0.5\linewidth}

\begin{tikzpicture}[scale=0.6]

\draw[thick,decoration={markings,mark=at position 0.6 with {\arrow[scale=1,>=stealth]{>}}},postaction={decorate}] (-1,0) node[left]{${2}$}--(-0.2,0);
\draw[thick,decoration={markings,mark=at position 0.6 with {\arrow[scale=1,>=stealth]{>}}},postaction={decorate}] (0,-1)node[below]{${1}$}--(0,0);
\draw[thick,decoration={markings,mark=at position 0.6 with {\arrow[scale=1,>=stealth]{>}}},postaction={decorate}] (0.2,0)--(1,0)node[right]{${4}$};
\draw[thick,decoration={markings,mark=at position 0.6 with {\arrow[scale=1,>=stealth]{>}}},postaction={decorate}] (0,0)--(0,1)node[above]{${3}$};

\node at (0,-2.3) {\small{$s$-channel}};

\draw[thick,rounded corners=5,decoration={markings,mark=at position 0.3 with {\arrow[scale=1,>=stealth]{>}}},decoration={markings,mark=at position 0.85 with {\arrow[scale=1,>=stealth]{>}}},postaction={decorate}] (0+4,-1) node[below]{${1}$}--(0+4,0)--(1+4,0)node[right]{${4}$};
\draw[thick,rounded corners=5,decoration={markings,mark=at position 0.3 with {\arrow[scale=1,>=stealth]{>}}},decoration={markings,mark=at position 0.85 with {\arrow[scale=1,>=stealth]{>}}},postaction={decorate}] (-1+4,0)node[left]{${2}$}--(0+4,0)--(0+4,1)node[above]{${3}$};

\node at (4,-2.3) {\small{$t$-channel}};

\draw[thick,rounded corners=5,decoration={markings,mark=at position 0.3 with {\arrow[scale=1,>=stealth]{>}}},decoration={markings,mark=at position 0.85 with {\arrow[scale=1,>=stealth]{<}}},postaction={decorate}] (0+8,-1) node[below]{${1}$}--(0+8,0)--(-1+8,0)node[left]{${2}$};
\draw[thick,rounded corners=5,decoration={markings,mark=at position 0.3 with {\arrow[scale=1,>=stealth]{<}}},decoration={markings,mark=at position 0.85 with {\arrow[scale=1,>=stealth]{>}}},postaction={decorate}] (1+8,0)node[right]{${4}$}--(0+8,0)--(0+8,1)node[above]{${3}$};

\node at (8,-2.3) {\small{ $u$-channel}};

%
%
%
\end{tikzpicture}

\caption{\label{fig:2qubitc}
$s$, $t$ and $u$ channels: non-orthogonal intertwiner basis\\}

\end{subfigure}

\caption{4-valent intertwiner maps as a 2-qubit gates.}
\label{fig:2qubit}
\end{figure}
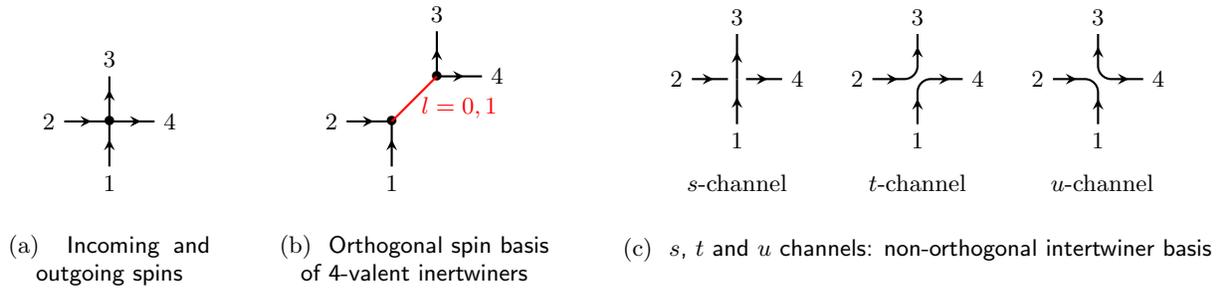

Let us focus on a single time slice to begin with.
To make things completely explicit and easily computable, we fix all the spins to $\f12$, that the incoming spins $j_{1}=j_{2}=\f12$, the outgoing spins $\tj_{1}=\tj_{2}=\f12$ and the transverse spins $k_{1}=k_{2}=\f12$.
Referring to a $\f12$ spin state as a qubit, the two intertwiners $I_{\alpha}$ and $I_{\beta}$ map a pair of qubits to another pair of qubits. They can be understood as 2-qubit gates:
\be
I_{v}:\,\cV_{\f12}\otimes \cV_{\f12}\rightarrow \cV_{\f12}\otimes \cV_{\f12}\,,
\ee
where we number the four spins around the node according to fig.\ref{fig:2qubita}, with incoming spins 1 and 2 and outgoing spins 3 and 4.
Unlike the usual setting of quantum circuit, we do not require these 2-qubit gates to be unitary but we require them to be $\SU(2)$-invariant. Such maps span a two-dimensional Hilbert space, whose usual orthonormal basis is given by the projections on the two allowed values for the recoupled spin $l=0$ and $l=1$ (see e.g. \cite{Livine:2006xc}), as drawn on fig.\ref{fig:4spinintertwiner}:
\be
\label{2spinrecoupling0}
P_{0}=|\Omega\ra\la \Omega|\,,\qquad
|\Omega\ra=\f1{\sqrt{2}}\,\big{(}|\uparrow\da\ra-|\da\ua\ra\big{)}\,,
\ee
\be
\label{2spinrecoupling1}
P_{1}=|+\ra\la+|+|0\ra\la0|+|-\ra\la -|\,,\qquad\textrm{with}\quad
\left|\begin{array}{lcl}
|+\ra&=&|\ua\ua\ra\\
|0\ra&=&(|\ua\da\ra+|\da\ua\ra)/\sqrt{2}\\
|-\ra&=&|\da\da\ra
\end{array}\right.
\,\,,
\ee
where we have used the standard notation with up and down arrows for the positive and negative magnetic moments, $|\ua\ra$ for $m=+\f12$ and $|\da\ra$ for $m=-\f12$. It is fairly straightforward to check that these are orthonormal projectors:
\be
P_{0}^{2}=P_{0}\,,\quad
P_{1}^{2}=P_{1}\,,\quad
P_{0}P_{1}=P_{1}P_{0}=0
\,.\nn
\ee

Instead of this basis, we choose to work with the non-orthonormal basis formed by the $s$-channel intertwiner $I^s$ and the $t$-channel intertwiner $I^{t}$. As used in \cite{Feller:2015yta,Dittrich:2017hnl}, the $s$-channel corresponds to the spins $1$ and $3$ recoupling to a vanishing spin, while $s$-channel corresponds to the spins $1$ and $4$ recoupling to a vanishing spin, as illustrated on fig.\ref{fig:2qubitc}. These are simply the identity map and the swap map:
\be
I^{s}|v\ra\otimes |w\ra=|v\ra\otimes |w\ra\,,\qquad
I^{t}|v\ra\otimes |w\ra=|w\ra\otimes |v\ra\,.
\ee
These two maps are obviously  $\SU(2)$-invariant. They are also unitary, unlike the projector basis maps.
Now, each intertwiner, at the vertices $\alpha$ and $\beta$, can be a superposition of these two basis states:
\be
I_{\alpha}=s_{\alpha}I^{s}+t_{\alpha}I^{t}\,,\qquad
I_{\beta}=s_{\beta}I^{s}+t_{\beta}I^{t}\,,
\ee
with a priori arbitrary complex coefficients $s_{v},t_{v}\in\C$. By simple diagrammatics, we see that one-slice transfer matrix is also a superposition of the identity and the swap maps:
\be
T:\cV_{j_{1}}\otimes\cV_{j_{2}}\longrightarrow\cV_{\tj_{1}}\otimes\cV_{\tj_{2}}\,,\qquad
T_{\{s_{v},t_{v}\}}=(2s_{\alpha}s_{\beta}+t_{\alpha}s_{\beta}+s_{\alpha}t_{\beta})\id_{2}+t_{\alpha}t_{\beta}\cS
\,,
\ee
where we took care of the factor 2 coming from tracing over the transverse loop. We wrote $\id_{2}$ and $\cS$ for the identity and swap maps instead of $I^{s}$ and $I^{t}$, although these are exactly the same maps, in order to clearly distinguish the intertwiner maps living at the spin network nodes from the overall transfer matrix and evolution map. Written as such, the transfer matrix clearly commutes with the $\SU(2)$ action, i.e. it is an intertwiner between the incoming and outgoing spins.

\medskip

To understand the spectrum of the transfer matrix $T$ and the ensuing dynamical selection of the total spin $J$, we can compute the action of $T$ directly in the recoupled basis of $\cH^{\pp}=\cV_{j_{1}}\otimes\cV_{j_{2}}$, as given above in (\ref{2spinrecoupling0}-\ref{2spinrecoupling1}):
\be
|\Omega\ra=|(\tfrac12,\tfrac12)\,J=0\ra=\tfrac1{\sqrt{2}}\,\big{(}|\uparrow\da\ra-|\da\ua\ra\big{)}
\,,\qquad
\left|\begin{array}{lclcl}
|+\ra&=&|(\f12,\f12)\,J=1,M=1\ra&=&|\ua\ua\ra\vspace*{1mm}\\
|0\ra&=&|(\f12,\f12)\,J=1,M=0\ra&=&\tfrac1{\sqrt{2}}\,(|\ua\da\ra+|\da\ua\ra)\vspace*{1mm}\\
|-\ra&=&|(\f12,\f12)\,J=1,M=-1\ra&=&|\da\da\ra
\end{array}\right.
\ee
Interestingly, both maps $\id_{2}$ and $\cS$ leave all those basis states invariant, except for the swap map acting on the 0-spin state, $\cS\,|\Omega\ra=\,-|\Omega\ra$. This means that $T$ is diagonalized by the recoupled basis with eigenvalues:
\be
\lambda_{0}=2s_{\alpha}s_{\beta}+t_{\alpha}s_{\beta}+s_{\alpha}t_{\beta}-t_{\alpha}t_{\beta}
\,,\quad\textrm{and}\quad
\lambda_{1}=2s_{\alpha}s_{\beta}+t_{\alpha}s_{\beta}+s_{\alpha}t_{\beta}+t_{\alpha}t_{\beta}
\,,
\ee
respectively for the recoupled spin subspaces $J=0$ and $J=1$.

It is clear that the eigenvalues of the transfer matrix are determined by the choice of intertwiners defining the boundary spin network and that they depend only on the value of the recoupled spin $J$. Since the eigenvalues for $J=0$ and $J=1$ are different, the asymptotic behavior at late time can enhance this difference.
For instance, let us choose totally homogeneous boundary conditions, with $S$ identical time slices and the same intertwiner $I_{v}=sI^{s}+tI^{t}$ at every vertex. The two eigenvalues read:
\be
\lambda_{0}=2s(s+t)-t^{2}
\,,\quad\textrm{and}\quad
\lambda_{1}=2s(s+t)+t^{2}
\,.
\ee
The largest eigenvalue in modulus will dominate at late time as the number of time slices $S$ grows large towards infinity. Choosing as an example $s=1$ and $t=0<\eps<1$, we get:
\be
\lambda_{0}=2+2\eps-\eps^{2}\quad <\quad 2+2\eps+\eps^{2}=\lambda_{1}
\,,
\ee
leading to the maximum recoupled spin $J=1$ will dominate the state at late time. Similarly, introducing a phase between the two components of the intertwiners with $s=1$ but $t=i\eps$ leads to the opposite behavior,
\be
|\lambda_{0}|=|2+\eps^{2}+2i\eps|\quad >\quad |2-\eps^{2}+2i\eps|=|\lambda_{1}|
\,,
\ee
with the dynamical selection of the $J=0$ sector at late time.

\subsection{The three spin$\f12$ propagator}

Let us compute the Ponzano-Regge evolution map in a slightly larger boundary Hilbert space, moving to the case with $N=3$ disk insertions, yet sticking to the simplification of spins fixed to $\f12$. This leads to the transfer matrix driving the evolution of 3 qubits along the cylinder.
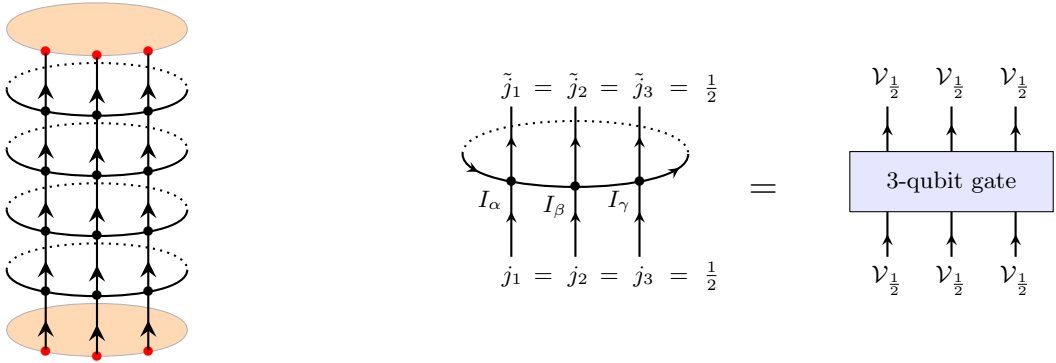
\begin{figure}[h!]

\begin{subfigure}{0.3\linewidth}
\begin{tikzpicture}[scale=.8]
\coordinate(ai) at (0,0);
\coordinate(af) at (0,5);
\coordinate(bi) at (3,0);
\coordinate(bf) at (3,5);
				
\draw[fill=orange, fill opacity=.3,opacity=.3](ai) to[in=-90,out=-90,looseness=.5]   node[pos=0.3,inner sep=0pt](ci){}node[pos=0.5,inner sep=0pt](di){}node[pos=0.7,inner sep=0pt](ei){} (bi)to[in=90,out=90,looseness=.5] (ai);

\draw[fill=orange, fill opacity=.3,opacity=.3](af) to[in=-90,out=-90,looseness=.5]   node[pos=0.3,inner sep=0pt](cf){}node[pos=0.5,inner sep=0pt](df){}node[pos=0.7,inner sep=0pt](ef){} (bf)to[in=90,out=90,looseness=.5] (af);
				
\foreach \i in {1,...,4}{
\draw[thick](0,\i) to[in=-90,out=-90,looseness=.5]  node[pos=0.3,inner sep=0pt]{$\bullet$}   node[pos=0.5,inner sep=0pt]{$\bullet$}  node[pos=0.7,inner sep=0pt]{$\bullet$}  (3,\i);
\draw[dotted,thick,in=90,out=90,looseness=.5] (0,\i) to (3,\i);
}
						

\node at (ci) {$\red{\bullet}$};
\node at (cf) {$\red{\bullet}$};
\node at (di) {$\red{\bullet}$};
\node at (df) {$\red{\bullet}$};
\node at (ei) {$\red{\bullet}$};
\node at (ef) {$\red{\bullet}$};
\draw[thick,decoration={markings,mark=at position 0.1 with {\arrow[scale=1.5,>=stealth]{>}}},decoration={markings,mark=at position 0.3 with {\arrow[scale=1.5,>=stealth]{>}}},decoration={markings,mark=at position 0.5 with {\arrow[scale=1.5,>=stealth]{>}}},decoration={markings,mark=at position 0.7 with {\arrow[scale=1.5,>=stealth]{>}}},decoration={markings,mark=at position 0.9 with {\arrow[scale=1.5,>=stealth]{>}}},postaction={decorate}] (ci)--(cf);
\draw[thick,decoration={markings,mark=at position 0.1 with {\arrow[scale=1.5,>=stealth]{>}}},decoration={markings,mark=at position 0.3 with {\arrow[scale=1.5,>=stealth]{>}}},decoration={markings,mark=at position 0.5 with {\arrow[scale=1.5,>=stealth]{>}}},decoration={markings,mark=at position 0.7 with {\arrow[scale=1.5,>=stealth]{>}}},decoration={markings,mark=at position 0.9 with {\arrow[scale=1.5,>=stealth]{>}}},postaction={decorate}] (di)--(df);
\draw[thick,decoration={markings,mark=at position 0.1 with {\arrow[scale=1.5,>=stealth]{>}}},decoration={markings,mark=at position 0.3 with {\arrow[scale=1.5,>=stealth]{>}}},decoration={markings,mark=at position 0.5 with {\arrow[scale=1.5,>=stealth]{>}}},decoration={markings,mark=at position 0.7 with {\arrow[scale=1.5,>=stealth]{>}}},decoration={markings,mark=at position 0.9 with {\arrow[scale=1.5,>=stealth]{>}}},postaction={decorate}] (ei)--(ef);
\end{tikzpicture}
\end{subfigure}
\hspace*{5mm}
\begin{subfigure}{0.6\linewidth}
\begin{tikzpicture}[scale=1]

\draw[thick,decoration={markings,mark=at position 0.1 with {\arrow[scale=1,>=stealth]{>}}},decoration={markings,mark=at position .93 with {\arrow[scale=1,>=stealth]{>}}},postaction={decorate}](0,0) to[in=-90,out=-90,looseness=.5]  node[pos=0.3,inner sep=0pt](a){$\bullet$} node[pos=0.5,inner sep=0pt](c){$\bullet$} node[pos=0.7,inner sep=0pt](b){$\bullet$}  (3,0);
\draw[dotted,thick,in=90,out=90,looseness=.5] (0,0) to (3,0);

\draw[thick,decoration={markings,mark=at position 0.3 with {\arrow[scale=1,>=stealth]{>}}},decoration={markings,mark=at position 0.8 with {\arrow[scale=1,>=stealth]{>}}},postaction={decorate}] ($(a)+(0,-1)$)--($(a)+(0,1)$);
\node[below left] at (a) {$I_{\alpha}$};
\draw[thick,decoration={markings,mark=at position 0.3 with {\arrow[scale=1,>=stealth]{>}}},decoration={markings,mark=at position 0.8 with {\arrow[scale=1,>=stealth]{>}}},postaction={decorate}] ($(b)+(0,-1)$)--($(b)+(0,1)$);
\node[below left] at (b) {$I_{\gamma}$};
\draw[thick,decoration={markings,mark=at position 0.3 with {\arrow[scale=1,>=stealth]{>}}},decoration={markings,mark=at position 0.8 with {\arrow[scale=1,>=stealth]{>}}},postaction={decorate}] ($.5*(a)+.5*(b)+(0,-1)$)--($.5*(a)+.5*(b)+(0,1)$);
\node[below left] at (c) {$I_{\beta}$};
\node at ($.5*(a)+.5*(b)+(0.47,-1.25)$) {$j_{1}\,=\,\,j_{2}\,=\,\,j_{3}\,\,=\,\,\f12$};
\node at ($.5*(a)+.5*(b)+(0.47,1.25)$) {$\tj_{1}\,=\,\,\tj_{2}\,=\,\,\tj_{3}\,\,=\,\,\f12$};
%

\node at (4,-.5) {\Large{=}};

\coordinate(x1) at  ($(a)+(5,-1)$);
\coordinate(y1) at  ($(a)+(5,1)$);
\coordinate(z1) at  ($(a)+(5,-0.4)$);
\coordinate(t1) at  ($(a)+(5,.4)$);
\coordinate(x2) at  ($(b)+(5,-1)$);
\coordinate(y2) at  ($(b)+(5,1)$);
\coordinate(z2) at  ($(b)+(5,-0.4)$);
\coordinate(t2) at  ($(b)+(5,.4)$);
\coordinate(z0) at  ($(z1)+(-.5,0)$);
\coordinate(t0) at  ($(t1)+(-.5,0)$);
\coordinate(z3) at  ($(z2)+(.5,0)$);
\coordinate(t3) at  ($(t2)+(.5,0)$);
\coordinate(x4) at  ($.5*(a)+.5*(b)+(5,-1)$);
\coordinate(y4) at  ($.5*(a)+.5*(b)+(5,1)$);
\coordinate(z4) at  ($.5*(a)+.5*(b)+(5,-0.4)$);
\coordinate(t4) at  ($.5*(a)+.5*(b)+(5,.4)$);

\draw[thick,decoration={markings,mark=at position 0.5 with {\arrow[scale=1,>=stealth]{>}}},postaction={decorate}] (x1)--(z1);
\draw[thick,decoration={markings,mark=at position 0.6 with {\arrow[scale=1,>=stealth]{>}}},postaction={decorate}] (t1)--(y1);
\draw[thick,decoration={markings,mark=at position 0.5 with {\arrow[scale=1,>=stealth]{>}}},postaction={decorate}] (x2)--(z2);
\draw[thick,decoration={markings,mark=at position 0.6 with {\arrow[scale=1,>=stealth]{>}}},postaction={decorate}] (t2)--(y2);
\draw[thick,decoration={markings,mark=at position 0.5 with {\arrow[scale=1,>=stealth]{>}}},postaction={decorate}] (x4)--(z4);
\draw[thick,decoration={markings,mark=at position 0.6 with {\arrow[scale=1,>=stealth]{>}}},postaction={decorate}] (t4)--(y4);
\draw[fill=blue, fill opacity=.1](z0)--(z3)--(t3)--(t0)--(z0);
\node at ($.5*(a)+.5*(b)+(5,0)$) {3-qubit gate};
\node[below] at (x1) {$\cV_{\f12}$};
\node[below] at (x2) {$\cV_{\f12}$};
\node[above] at (y1) {$\cV_{\f12}$};
\node[above] at (y2) {$\cV_{\f12}$};
\node[below] at (x4) {$\cV_{\f12}$};
\node[above] at (y4) {$\cV_{\f12}$};

\end{tikzpicture}
\end{subfigure}

\caption{Three spin states evolving across a square lattice on the 2d cylinder: time slices as 3-spin gates}
\label{fig:N3evolution}
\end{figure}

As drawn on fig.\ref{fig:N3evolution}, one time slice consists in the three qubits getting recoupled by three intertwiners linked by a common tranverse spin looping around the cylinder. Having also fixed the transverse spin to $\f12$, the spin network data on a single time slice is given by the choice of the three intertwiners living on the lattice nodes:
\be
I_{v}=s_{v}I^{s}+t_{v}I^{t}\,,\qquad\textrm{for}\quad v=\alpha,\beta,\gamma\,.
\ee
Following the simple diagrammatics of the $s$ and $t$ channels for each intertwiner leads to the following transfer matrix:
\be
T:\cV_{j_{1}}\otimes\cV_{j_{2}}\otimes\cV_{j_{3}}\longrightarrow\cV_{\tj_{1}}\otimes\cV_{\tj_{2}}\otimes\cV_{\tj_{3}}\,,
\nn
\ee
\be
T_{\{s_{v},t_{v}\}}
=
(2s_{\alpha}s_{\beta}s_{\gamma}+t_{\alpha}s_{\beta}s_{\gamma}+s_{\alpha}t_{\beta}s_{\gamma}+s_{\alpha}s_{\beta}t_{\gamma})\id_{3}
+s_{\alpha}t_{\beta}t_{\gamma}\cP_{23}+t_{\alpha}s_{\beta}t_{\gamma}\cP_{13}+t_{\alpha}t_{\beta}s_{\gamma}\cP_{12}
+t_{\alpha}t_{\beta}t_{\gamma}\cC_{(231)}
\,,
\ee
where $\cP_{ab}$ is the permutation between the qubits $a$ and $b$ and $\cC_{(231)}$ is the clockwise cycle:
\be
\cP_{23}|u\ra\otimes|v\ra\otimes |w\ra=|u\ra\otimes|w\ra\otimes |v\ra
\,,\quad
\cC_{(231)}|u\ra\otimes|v\ra\otimes |w\ra=|v\ra\otimes|w\ra\otimes |u\ra
\,.
\ee
Let us look at the action of this evolution map on the recoupled basis state for 3 qubits. We construct the recoupled basis by first recoupling the first two qubits and then recoupling that pair to the third qubit:
\be
(\cV_{\f12})^{\otimes 3}
=
(\cV_{\f12}\otimes \cV_{\f12})\otimes\cV_{\f12}
=
(\cV_{0}\oplus \cV_{1})\otimes\cV_{\f12}
=
\cV_{\f12}\oplus (\cV_{\f12}\oplus \cV_{\f32})
=
2\cV_{\f12}\oplus \cV_{\f32}
\,.
\ee
The resulting basis states are:
\be
|0,\ua\ra\equiv|J_{12}=0,J=\tfrac12,M=+\tfrac12\ra
=
\tfrac1{\sqrt{2}}\,\big{(}|\ua\da\ua\ra-|\da\ua\ua\ra\big{)}
\,,\quad
|0,\da\ra\equiv|J_{12}=0,J=\tfrac12,M=-\tfrac12\ra
=
\tfrac1{\sqrt{2}}\,\big{(}|\ua\da\da\ra-|\da\ua\da\ra\big{)}\,,
\nn
\ee
\be
|1,\ua\ra\equiv|J_{12}=1,J=\tfrac12,M=+\tfrac12\ra
=
\sqrt{\tfrac23}|\ua\ua\da\ra
-\tfrac1{\sqrt6}\Big{[}
|\ua\da\ua\ra+|\da\ua\ua\ra
\Big{]}\,,\nn
\ee
\be
|1,\da\ra\equiv|J_{12}=1,J=\tfrac12,M=-\tfrac12\ra
=
\tfrac1{\sqrt6}\Big{[}
|\da\ua\da\ra+|\ua\da\da\ra
\Big{]}
-
\sqrt{\tfrac23}|\da\da\ua\ra
\,,\nn
\ee
\be
\begin{array}{lclcl}
|+\tfrac32\ra
&=&
|J_{12}=1,J=\tfrac32,M=+\tfrac32\ra
&=&
|\ua\ua\ua\ra\,, 
\vspace*{1mm}\\
|+\tfrac12\ra
&=&
|J_{12}=1,J=\tfrac32,M=+\tfrac12\ra
&=&
\tfrac1{\sqrt{3}}\Big{[}
|\da\ua\ua\ra+|\ua\da\ua\ra+|\ua\ua\da\ra
\Big{]}\,, 
\vspace*{1mm}\\
|-\tfrac12\ra
&=&
|J_{12}=1,J=\tfrac32,M=-\tfrac12\ra
&=&
\tfrac1{\sqrt{3}}\Big{[}
|\da\da\ua\ra+|\da\ua\da\ra+|\ua\da\da\ra
\Big{]}\,, 
\vspace*{1mm}\\
|-\tfrac32\ra
&=&
|J_{12}=1,J=\tfrac32,M=-\tfrac32\ra
&=&
|\da\da\da\ra\,.
\end{array}
\ee
The interesting feature of this $N$=3 qubit case compared to the previous $N$=2 qubit case is the non-trivial multiplicity for the recoupled spin $J=\tfrac12$. This opens the door to a non-trivial dynamics between the two recoupled subspaces $\cV_{\f12}$: will they carry the same or different eigenvalue(s) of the transfer matrix $T$?

\medskip

Let us compute the action of the relevant permutations on these basis states. First, all the states for the maximal recoupled spin $J=\tfrac32$ are left invariant under the permutations $\cP_{ab}$ and $\cC_{(231)}$. The states for $J=\tfrac12$ have a non-trivial behavior but are mapped onto linear combinations of states with same values of $J$ and $M$ (as expected due to the $\SU(2)$-invariance property):
\be
\left|\begin{array}{lcl}
\cP_{12}|0,\ua\ra
&=&
-|0,\ua\ra
\,,\\
\cP_{23}|0,\ua\ra
&=&
\tfrac{\sqrt{3}}2|1,\ua\ra+\f12|0,\ua\ra
\,,\\
\cP_{13}|0,\ua\ra
&=&
-\tfrac{\sqrt{3}}2|1,\ua\ra+\f12|0,\ua\ra
\,,\\
\cC_{(231)}|0,\ua\ra
&=&
\tfrac{\sqrt{3}}2|1,\ua\ra-\f12|0,\ua\ra
\,,
\end{array}\right.
\qquad
\left|\begin{array}{lcl}
\cP_{12}|0,\da\ra
&=&
-|0,\da\ra
\,,\\
\cP_{23}|0,\da\ra
&=&
\tfrac{\sqrt{3}}2|1,\da\ra+\f12|0,\da\ra
\,,\\
\cP_{13}|0,\da\ra
&=&
-\tfrac{\sqrt{3}}2|1,\da\ra+\f12|0,\da\ra
\,,\\
\cC_{(231)}|0,\da\ra
&=&
\tfrac{\sqrt{3}}2|1,\da\ra-\f12|0,\da\ra
\,,
\end{array}\right.
\ee
\be
\left|\begin{array}{lcl}
\cP_{12}|1,\ua\ra
&=&
|1,\ua\ra
\,,\\
\cP_{23}|1,\ua\ra
&=&
-\f12|1,\ua\ra+\tfrac{\sqrt{3}}2|0,\ua\ra
\,,\\
\cP_{13}|1,\ua\ra
&=&
-\f12|1,\ua\ra-\tfrac{\sqrt{3}}2|0,\ua\ra
\,,\\
\cC_{(231)}|1,\ua\ra
&=&
-\f12|1,\ua\ra-\tfrac{\sqrt{3}}2|0,\ua\ra
\,,
\end{array}\right.
\qquad
\left|\begin{array}{lcl}
\cP_{12}|1,\da\ra
&=&
|1,\da\ra
\,,\\
\cP_{23}|1,\da\ra
&=&
-\f12|1,\da\ra+\tfrac{\sqrt{3}}2|0,\da\ra
\,,\\
\cP_{13}|1,\da\ra
&=&
-\f12|1,\da\ra-\tfrac{\sqrt{3}}2|0,\da\ra
\,,\\
\cC_{(231)}|1,\da\ra
&=&
-\f12|1,\da\ra-\tfrac{\sqrt{3}}2|0,\da\ra
\,.
\end{array}\right.
\ee
Let us study in details the homogeneous case where the intertwiner is the same at the three nodes, $I_{\alpha}=I_{\beta}=I_{\gamma}$. Thus, assuming that the coefficients $s_{v}$ and $t_{v}$ do not depend on the vertex, the expression of the transfer matrix simplifies a little bit:
\be
T=s^{2}(2s+3t)\id_{3}+st^{2}(\cP_{23}+\cP_{13}+\cP_{12})+t^{3}\cC_{(231)}\,.
\ee
The action of this map on states with maximal recoupled spin $J=\f32$ is proportional to the identity since those states are invariant under permutations:
\be
T|J=\tfrac32,M\ra
=
(2s^{3}+3s^{2}t+3st^{2}+t^{3})\,|J=\tfrac32,M\ra
\,,
\ee
where we have purposefully omitted the obvious label $J_{12}=1$.
On the other hand, for the $J=\f12$ sector, the permutations operators $\cP_{ab}$ and $\cC_{(231)}$ act non-trivially. Since they leave the label $J,M$ invariant, they are understood to act solely on the $J_{12}$ label. Thus, omitting the label $J=\f12$ and the $M=\pm\f12$, the permutation operators act as:
\be
(\cP_{23}+\cP_{13}+\cP_{12})|J_{12}\ra=0\,,\qquad
\left|\begin{array}{lcl}
\cC_{(231)}|J_{12}=0\ra
&=&
-\f12|0\ra+\tfrac{\sqrt{3}}2|1\ra\,,
\\
\cC_{(231)}|J_{12}=1\ra
&=&
-\tfrac{\sqrt{3}}2|0\ra-\f12|1\ra\,.
\end{array}\right.
\ee
The action of the circular permutation is easily diagonalized by considering superpositions of the two subspaces with $J_{12}=0$ and $J_{12}=1$:
\be
\cC_{(231)}\,(|0\ra\pm i|1\ra)
=
-e^{\pm i \f\pi3}
(|0\ra\pm i|1\ra)\,.
\ee
This leads to two eigenvalues for the transfer matrix in the $J=\f12$ sector:
\be
T\,|J=\f12,M\ra_{\pm}
=
\Big{[}2s^{3}+3s^{2}t-t^{2}e^{\pm i \f\pi3}\Big{]}
\,|J=\f12,M\ra_{\pm}
\,,
\ee
\be
\textrm{with}\quad
|J=\f12,M\ra_{\pm}=|J=\f12,M,J_{12}=0\ra\pm i |J=\f12,M,J_{12}=1\ra
\,.
\nn
\ee
The transfer matrix therefore admits three eigenvalues, one for the $J=\f32$ sector and two for the $J=\f12$ sector. The largest eigenvalue will dominate the transition amplitude in the limit of an infinite number of time slices $S\rightarrow+ \infty$. This limit can be understood either as a late time regime or as the infinite refinement of the time direction. 

\begin{figure}[h!]

\begin{subfigure}[t]{0.3\linewidth}
\includegraphics[width=50mm]{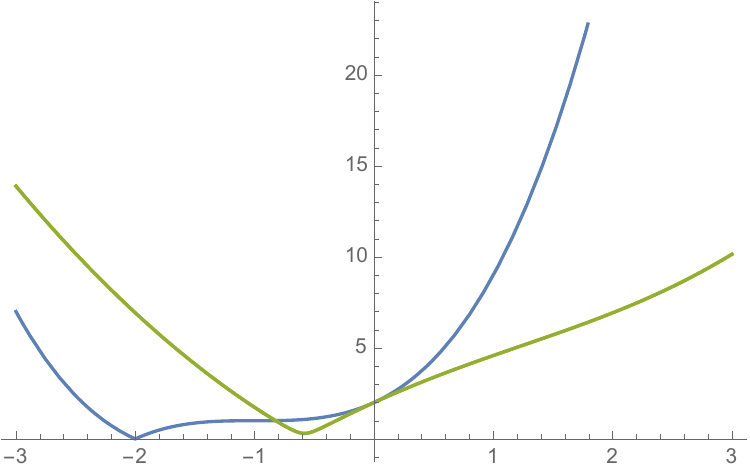}
\caption{ $t\in\R$:  the two eigenvalues of the $J$=$\f12$ have same modulus, $|\lambda_{\f12}^+|=|\lambda_{\f12}^-|$.}

\end{subfigure}
\hspace*{2mm}
\begin{subfigure}[t]{0.32\linewidth}
\includegraphics[width=50mm]{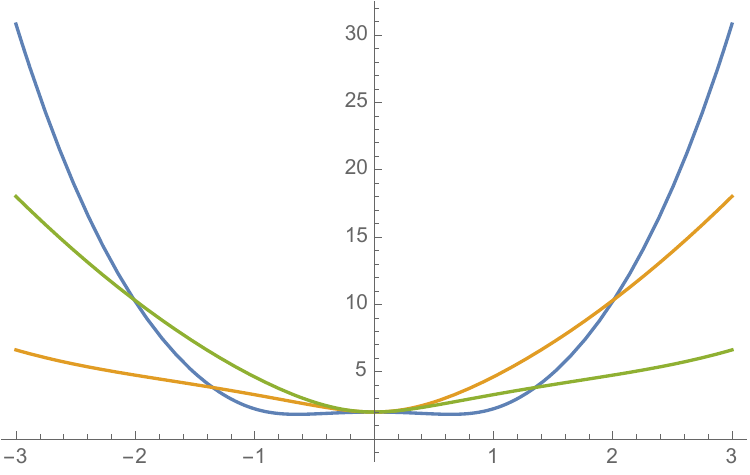}
\caption{ $t\in i\R$:  the modulus of two eigenvalues of the $J$=$\f12$ are equal up to the sign switch $t\leftrightarrow -t$.}
\end{subfigure}
\hspace*{2mm}
\begin{subfigure}[t]{0.3\linewidth}
\includegraphics[width=50mm]{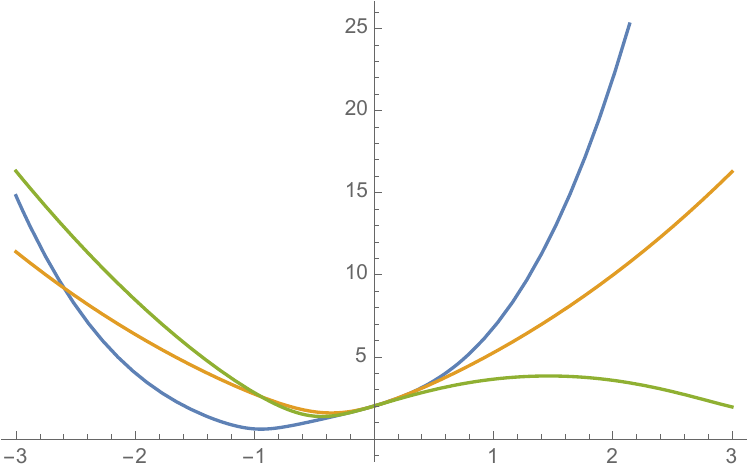}
\caption{$t\in e^{i\f\pi4}\R$: the sector $J$=$\f32$ dominates for positive values, while the sector $J$=$\f12$ dominates for small negative values.}
\end{subfigure}

\caption{Modulus of the three eigenvalues of the transfer matrix $T$ describing the time evolution of $N$=3 qubis for a single time slice for the intertwiner parameters $s=1$ and varying $t\in\C$: the recoupled spin sector with the highest eigenvalue will dominate the dynamics in the limit of an infinite number of time slices $S\rightarrow+\infty$. This thermodynamic limit can be understood as a late time limit or an infinite refinement limit of the time coordinate. The eigenvalue $|\lambda_{\f32}|$ for the $J$=$\f32$ subspace is plotted in {\blue blue} while the two eigenvalues $|\lambda_{\f12}^\pm|$ for the $J$=$\f12$ sector are plotted respectively in {\green green} and {\orange orange}.}
\label{fig:plotN3}
\end{figure}

Let us look more explicitly at how the three eigenvalues depend on the choice of intertwiner. Setting $s=1$, we let $t$ vary. The eigenvalues read:
\be
\lambda_{\f32}=2+3t+3t^{2}+t^{3}
\,,\qquad
\lambda_{\f12}^\pm=2+3t-t^{2}e^{\pm i \f\pi3}
\,.
\ee
As the intertwiner parameter $t$ varies in the complex plane, the dominant eigenvalue changes, as one can see on the plots on fig.\ref{fig:plotN3}.

\medskip

This example illustrates generic features of the transfer matrix defined by the Ponzano-Regge model on a solid cylinder with a boundary spin network on a square lattice on the boundary cylinder:
\begin{itemize}
\item The transfer matrix leaves the recoupled spin data $J,M$ invariant, but have a good action on the multiplicity space $\cN_{J}$;
\item It can admit different eigenvalues within a $\cN_{J}$ subspace;
\item the eigenvalues do depend on $J$ but do not depend on $M$.
\end{itemize}

\medskip

One could now move on to a higher number $N$ of open edges on the disk boundary, which would mean to studying the (time) evolution of $N$ qubits as they go through the boundary spin network on the cylinder. The evolution map $\cT$ would consist in the iterated transfer matrix $T$. As in the examples above for two and three spin states on the corner, this transfer matrix would be expressed as a linear combination of permutations of the $N$ qubits, as shown in \cite{Dittrich:2017hnl}. Similarly, recoupled spin states can also be expressed in terms of representations of the permutation group, e.g. \cite{Livine:2005mw}. 
This could possibly lead to a systematic calculation of the transfer matrix eigenvalues in terms of permutations, leading to an explicit computation of the propagator resulting from the Ponzano-Regge model and a general understanding of the subspaces dominating  the dynamics in the infinite refinement limit of time $S\rightarrow+ \infty$.


\section*{Outlook \& Conclusion}

We have looked into the transition amplitudes in the Ponzano-Regge model for 3d quantum gravity for the boundary state of 2d disks evolving along a cylinder the time direction, which we refer to, in short, as the Ponzano-Regge propagator. A canonical boundary state for a 2d disk with $N$ holonomy insertions consists in the tensor product of $N$ spin states, one for each boundary insertion. The propagator then depends on the boundary spin network on the 2d cylinder interpolating between the initial and final disks, and is simply given by the evaluation of that boundary spin network.

The main feature of the resulting transition amplitudes is that they preserve  the total recoupled spin of the initial state. We work this out explicitly for boundary spin networks living on a square lattice on the 2d cylinder. Then one slice of the spin network defines a transfer matrix and the propagator is that transfer matrix taken to the power of the number of slices making the cylinder. Each value of the total recoupled spin corresponds to a different eigenvalue of the transfer matrix. Thus the highest eigenvalue, dominating the large time limit (as the number of slices grows large), will select a dominant a total recoupled spin, depending on the details of the considered spin network. In the case that the dominant spin is the 0 spin, this provides a dynamical mechanism for the emergence of an effective $\SU(2)$ symmetry.

To illustrate this, we computed the transfer matrix and its spectrum for the case of $N=2$ spins, both fixed to the elementary geometry excitation $j_{1}=j_{2}=\f12$, and for the case of $N=3$ spins also fixed to the lowest non-trivial value $j_{1}=j_{2}=j_{3}=\f12$. This amounts to studying the evolution of respectively two and three boundary qubits dictated by 3d quantum gravity. The transfer matrix for the square lattice can then be understood as respectively a two- or three- qubit gate involving an auxiliary qubit travelling along the transverse link and depending on the intertwiners dressing the boundary spin network.

\medskip

Now that the frame for the Ponzano-Regge propagator is set, the natural question is how it scales as the number of insertions $N$ (i.e. the boundary size in the space direction) grows and the number of slices (i.e. the time interval) $S$ grows. Taking $N$ and $S$ to infinity yields two different types of thermodynamical limits, which do not raise the same issues.

On the one hand, increasing the number of boundary insertions $N$ increases the complexity of the transfer matrix. It was shown in \cite{Dittrich:2017hnl} that the transfer matrix  for a boundary spin network based on a square lattice with elementary spins $\f12$ on all its links can be understood in terms of the simplest 2d integrable models: the 6-vertex model and equivalently the Heisenberg spin chain. The goal would be to import the result from integrable models, which extract the spectrum of the transfer matrix using the Bethe ansatz, and use those powerful methods to describe the evolution dictated by the Ponzano-Regge model.

%
On the other hand, the obvious question when considering the time direction is whether the evolution is unitary or not. In the cases that we studied, the spin network states for two and three qubits clearly didn't lead to a unitary transfer matrix.
This non-unitary evolution\footnotemark{} is to be expected since we are not studying a closed isolated system but an open quantum system with boundary.
The question should then be reformulated as how to identify if the boundary conditions, formalized at the quantum level as the spin network on the time-like boundary cylinder, allow for incoming or/and outgoing flux across the boundary. If the boundary conditions are rigid enough to freeze all flux exchange with the exterior of the cylinder and isolate the system, it would then make sense to require the unitary of the evolution. Following the reverse logic, one could interpret the unitarity of the evolution as the signature of an isolated system and attempt to relate the distance from unitarity to the flux of information across the boundary.
\footnotetext{
Another source of non-unitarity is also the discreteness of the time direction, but it seems that this is likely to translate into an intrinsic decoherence and map pure states into mixed states (see e.g. \cite{Milburn:2003zj}). Moreover, using coherent spin network states on the boundary still leads to non-unitary evolution maps \cite{Goeller:2019zpz} and unitarity seems to be recovered only in a very specific fine-tuned refinement limit as the number of time slices is sent to infinity, $S\rightarrow+\infty$.
}


\medskip

Beyond the exploration of possible extensions of the present results in the realm of 3d quantum gravity, our work further opens possible doors to applications beyond the the Ponzano-Regge model:

\begin{itemize}

\item {\bf Quantum circuitry:}

The boundary spin network evaluation, when restricted to spins $\f12$ on every link, defines a quantum circuit between the qubits on the initial disk's boundary and the qubits on the final disk's boundary. Considering spin networks on a square lattice leads to a very simple quantum circuit architecture for $N$ incoming qubits and stackable slices of $N$ 2-qubit gates. Those two qubits gates are $\SU(2)$-invariant gates, which be realized as superpositions of the identity map and the swap map. Then each slice couples the $N$ incoming qubits with one auxiliary qubit running transversally to the circuit. The lattice structure leads to an easily scalable quantum circuit, which could be used for quantum simulations of 3d quantum gravity sectors.

Those circuits could be also used, without link to 3d quantum gravity on the Ponzano-Regge model, as channels allowing to select specific $N$-qubit subspaces with fixed total recoupled spin.

\item {\bf Interplay with condensed matter \& experimental implementation:}

Furthermore, the lattice models dressed with spins $\f12$ on every link are understood to be equivalent to the  6-vertex model, itself equivalent to a 2d Ising model, solvable by Bethe ansatz \cite{Dittrich:2017hnl}. Not only this means that we could/should study and compute exactly the Ponzano-Regge propagator using integrable model methods, but, on more practical grounds, we could use physical realizations of the 2d Ising model to run actual experimental simulations of this 3d quantum gravity propagator along the cylinder.

\item {\bf Propagation in 3+1-d loop quantum gravity:}

%
Another range of possible applications comes from moving up from thee space-time dimensions to four space-time dimensions.
Let us keep in mind that 4d gravity is very different from 3d gravity: pure gravity in three space-time dimensions is a topological theory with non local degrees of freedom, which leads to a straightforward holography with a clear bulk-boundary relation, while 4d gravity has local degrees of freedom with non-trivial  bulk dynamics.

Loop quantum gravity in 3+1-dimensions also has $\SU(2)$ spin networks as its kinematical states (if working in the time gauge reducing the local Lorentz gauge symmetry to a local $\SU(2)$ gauge invariance and assuming a real value for the Immirzi parameter). These are the same spin networks as used in the Ponzano-Regge models. The difference comes in their interpretation as 3d space geometries instead of 2d geometries. The cylindric ansatz introduced in the present work would become relevant when looking at the propagation of ``closure defects'', which represent geometrical defects of torsion and curvature (see e.g. previous work on coarse-graining loop quantum gravity \cite{Livine:2013gna,Charles:2016xwc,Delcamp:2016lux, Livine:2019cvi}) and are thus interpretable as matter insertions. It would then represent waves of geometry propagating across the spin network.
Let us insist that this would be engineering kinematical states of the 3+1-d theory, on which one would still need to study the imposition of the Hamiltonian constraints encoding the dynamics of 4d gravity.

Pushing this logic further, one could also imagine a wave of geometry as an upgraded ansatz of a 2d grid of qubits propagating across a 3d cubic lattice. One would then attempt to generalize the algebraic structures derived here: what are the conserved observables commuting with the transfer matrix? how would  the propagation across the 3d lattice for a homogeneous choice of 6-valent intertwiners look like?
Once again, these 3d states of geometry with waves would be kinematical states and one would still require to investigate their evolution along the time direction (computing the corresponding spinfoam transition amplitudes) and the imposition of the quantum Einstein equations.

\end{itemize}

All these offers exciting prospects for the simple setting of the Ponzano-Regge propagator for 3d quantum gravity.


\appendix


\nocite{*}

\bibliographystyle{bib-style}
\bibliography{PR}

\end{document}